\documentclass[a4paper,fleqn,usenatbib]{mnras}

\usepackage{mathptmx}

\usepackage[T1]{fontenc}
\usepackage{ae,aecompl}
\usepackage[english]{babel}

\usepackage{hyperref}  
\usepackage[utf8]{inputenc}
\usepackage{booktabs, caption, makecell}

\usepackage{threeparttable}

\usepackage{graphicx}	
\usepackage{amsmath}	
\usepackage{amssymb}	

\usepackage{longtable}
\usepackage{lipsum}
\usepackage{txfonts}
\usepackage{threeparttable}
\usepackage[bottom]{footmisc}

\title[New Insights into Time Series Analysis III]{\centering New Insights into Time Series Analysis III \\ Setting constraints on period analysis}

\author[C. E. Ferreira Lopes et al.]{
C. E. Ferreira Lopes$^{1}$\thanks{E-mail: ferreiralopes1011@gmail.com (KTS)}
N. J. G.~Cross$^2$ 
F.  Jablonski$^{1}$
\\
$^{1}$National Institute For Space Research (INPE/MCTI), Av. dos Astronautas, 1758 – São José dos Campos – SP, 12227-010, Brazil  \\
$^{2}$SUPA (Scottish Universities Physics Alliance) Wide-Field Astronomy Unit, Institute for Astronomy, School of Physics and Astronomy,\\ University of Edinburgh, Royal Observatory, Blackford Hill, Edinburgh EH9 3HJ, UK
}

\date{Accepted XXX. Received YYY; in original form ZZZ}

\pubyear{2018}

\begin{document}
\label{firstpage}
\pagerange{\pageref{firstpage}--\pageref{lastpage}}
\maketitle

\begin{abstract}
E-science of photometric data requires
automatic procedures and a precise recognition of periodic patterns to perform 
science as well as possible on large data. Analytical
equations that enable us to set the best constraints to properly reduce
processing time and hence optimize signal searches play a crucial role in this
matter. These are increasingly important because the production of
unbiased samples from variability indices and statistical parameters has not
been achievable so far. We discuss the constraints used in periodic signals
detection methods as well as the uncertainties in the estimation of periods and
amplitudes. The frequency resolution 
necessary to investigate a time series is assessed with a new approach that 
estimates the necessary sampling resolution from shifts on the phase diagrams 
for successive frequency grid points.We demonstrate the underlying meaning of the
oversampling factor. We reassess the frequency resolutions required to
find the variability periods of $EA$ stars and use the new resolutions to
analyse a small sample of $EA_{up}$ Catalina stars, i.e. $EA$ stars previously 
classified as having insufficient number of observations at the eclipses.
As a result, the variability periods of four $EA_{up}$ stars were determined.
Moreover, we have a new approach to estimate the 
amplitude and period variations. 
From these estimations information about the intrinsic variations of the
sources are obtained. For a complete characterization of the light
curve signal the period uncertainty and period variation must be determined.
Constraints on periodic signal searches were analysed and delimited.
\end{abstract}




\section{Introduction}\label{introduction}	

Some time series are stochastic (or random) in the sense that they do not
contain underlying information other than noise. The analysis of large
databases requires automatic and efficient classifiers to provide the
identification of genuine features. This is crucial to reduce the
number of misclassifications, to narrow the boundaries between classes,
to provide better training sets, as well as to diminish the total processing 
time \citep[][]{Eyer-2006}. Large volumes of data containing potentially 
interesting scientific results are left unexplored or have their analysis 
delayed due to the current limited inventory of tools which are unable
to produce clean samples, despite big efforts having been undertaken. In fact, 
we risk underusing a large part of these data.
In order to improve the efficiency of variability indices, we propose 
discriminating variable sources in correlated and non-correlated data.
The correlated data have several measurements close in time, from
which accurate correlated indices are computed. On the other hand, the
non correlated data are those sources having too few measurements 
close in time and so they must be analysed using statistical parameters only.
The use of correlated and non-correlated indices \citep[see Sect.
4.3 in][]{FerreiraLopes-2016papI}, produces a substantially smaller number of 
time series which have to be further analysed. However, the resulting selection
is still three or more times larger than just the well-defined signals,
according to \citet[][]{FerreiraLopes-2016papI,FerreiraLopes-2017papII}.
This means that the set of preliminary selection criteria is unable to produce 
samples comprised only of variable stars, and so, it would be desirable
that the following steps of signal searching methods would provide
reliable identifications and accurate estimates of periods
(frequencies) and amplitudes, even in cases where the preliminary analysis
failed to give a confident indicator that the signal was truly variable
and not just a noisy time series.
Indeed, $\sim75\%$ of parameters used to characterize light curves are derived 
from the folded light curve using the variability periods
\citep[][]{Richards-2011}. This led to a $\sim 11\%$ misclassification
rate for non-eclipsing variable stars, for instance, \cite{Dubath-2011}.
Nowadays, reliable samples, i.e. samples composed only of variable stars, are 
increasingly becoming more important than complete samples, i.e samples having
all variable stars but also having a large number of misclassified non-variable 
stars, since visual inspection of all sources is unfeasible.
Therefore, an approach that allows us to get unbiased samples having correct
periods is mandatory to return quicker scientific results. Therefore, as a
continuation of the ``New Insights into Time Series Analysis'' project, the 
frequency finding methods are reviewed and improved.

The periodic signals finding methods can be separated into three
main groups if we consider how each component of the
figure of merit in the frequency grid is computed:
$M_S$ - each epoch provides a single term; $M_P$ - each term is
computed using a pair of epochs; $M_B$ - each term is computed by binning the
phase diagram. The Lomb Scargle and its generalization (LS - 
\citealt[][]{Lomb-1976,Scargle-1982} and GLS - \citealt[][]{Zechmeister-2009}) 
belong to the $M_S$ group. Each epoch is regarded as a single power spectrum
term and the periodogram is equivalent to a least squares fit of the folded
data at each frequency by a sine wave. Indeed, Fourier methods and their
branches are the simplest methods belong to the $M_S$ group. On the
other hand, the string length method \citep[STR - ][]{Dworetsky-1983} and the analysis of 
variance \citep[AOV and AOVMHW - ][]{Schwarzenberg-Czerny-1989,
Schwarzenberg-Czerny-1996} belong to the $M_P$ group. However, they follow
different approaches since the STR power spectrum is computed using pairs of
epochs in the phase diagram while AOV pairs epochs in time. Phase
dispersion minimization \citep[PDM and PDM2 - ][]{Stellingwerf-1978,
Stellingwerf-2011}, conditional entropy \citep[CE - ][]{Graham-2013}, 
supersmoother \citep[SS - ][]{Reimann-1994}, and correntropy kernel periodogram
\citep[CKP - ][]{Huijse-2012} belong to the $M_B$ group, where the power
spectrum is computed by binning the phase diagram. Wavelet analysis also has
been used to study time series  \citep[e.g.][]{Foster-1996,Bravo-2014}. However
it is more suitable to study the evolution of a signal over time and it
requires continuous observation. Currently, these are the main frequency finding
methods but there are many others \citep[e.g.][]{Huijse-2011,Kato-2012}.

The efficiency of the frequency 
finding methods has been tested in the last few years 
\citep[e.g.][]{Heck-1985,Swingler-1989,Schwarzenberg-1999,
Distefano-2012, Graham-2013}. Usually, the authors analyze the sensitivity as
well as the fraction of true periods recovered within a defined accuracy limit. 
Indeed, research using real data, including for instance irregular sampling, 
gaps, outliers, and errors, may provide more reliable results. Currently, the
most complete of these studies was performed by \citet[][]{Graham-2013}. The 
authors analysed 11 different methods using light curves of $78$ variable star 
types. The conditional entropy-based algorithm is the most optimal in terms of
completeness and speed according to the authors. However, most frequency
finding methods have a selection effect for the identification of weak
periodic signals \citet[][]{deJager-1989,Schwarzenberg-Czerny-1999}. Therefore 
a combination of all methods could potentially increase the recover rate close
to $100\%$ according to \citet[][]{Dubath-2011}. However, which method leads to
the correct period for a specific light curve within an automated strategy is an
open question. Moreover, the main frequencies computed by different
methods can be dissimilar and so two questions must be answered to determine the best way to
analyse a time series, i.e. how many frequency finding methods should be combined and how
to work out which period should be chosen when two or more methods
provide different results?

The frequency finding methods adopt, as the true
frequency ($\mathrm{f_{true}}$), one that defines the main periodic variation
displayed by the time series, based on a minimum or maximum of the quantity
being tested. However, the main frequency can be a harmonic of
$\mathrm{f_{true}}$ or related to an instrumental or spurious variation. It
means that only using the period finding method is not enough to set a reliable
period and so additional analyses are required. For example, the harmonic fits
can be used to set models and, using the $\chi^2$ distribution, establish
$\mathrm{f_{true}}$ and its reliability \citep[][]{FerreiraLopes-2015wfcam}.
However, what $\chi^2$ threshold, below which a time series may be considered
to have a reliable signal, and if the $\chi^2$ alone is enough to do that are
open questions. Theoretically, any signal having an amplitude greater than
the noise could be detected using a suitable method. The false alarm 
probability (\textit{FAP}) has been used to determine the typical power 
spectrum values of the noise and to discard variability due to noise alone.
However, sources lacking a periodic signal, such as aperiodic variations and
pulses, will also be discarded using this approach. 
All the period finding methods that depend on the phase diagram are unfit
to detect these signals because no frequency will return a smooth phase diagram.
Therefore, classifying a time series as noisy requires an investigation of
all signal types. On the other hand, correlated noise, seasonal variations, the
cadence or the phase coverage can lead to power values above the FAP indicating 
an applicability limit to using this approach to determine the reliability of
selections. Indeed, a large number of Monte-Carlo simulations are usually 
performed to determine the FAP and hence the running time required should also
be taken into account. Then our required list of improvements towards an
efficient automation of the variability analysis should include:
how to use the current period finding methods to determine the reliability of
signals? How to discriminate aperiodic from stochastic variations? Is it
possible to study all variation types using a single approach or are different
strategies required for different purposes? How to provide a standard cutoff
to determine reliable signals?

Currently, any frequency finding 
method is able to compute $\mathrm{f_{true}}$ using a single computation.
Therefore, the determination of $\mathrm{f_{true}}$ is performed after 
computing several times a function that is susceptible to the
smoothness of the phase diagrams depending on the method that is used. The
phase values are given by;

\begin{equation}
     \mathrm{\phi_{i} =  t_{i}\times f_{test} - {\rm \mathrm{\textit{G}}} \left[ t_{i}\times \mathrm{f_{test}} \right]},
   \label{eqphi}
\end{equation}

\noindent  where $t_{i}$ is the time, $\mathrm{f_{test}}$ is a test frequency,
and the function $\mathrm{\textit{G}}$ returns the integer part of $ t_{i}\times
\mathrm{f_{test}}$. This equation provides an interval of values ranging $0
\leq \Phi \leq 1$ where $\mathrm{f_{true}}$ is the frequency which returns the
smoothest phase diagram. The minimum  ($\mathrm{f_{min}}$) and maximum 
($\mathrm{f_{max}}$) frequency as well as the resolution ($\Delta \mathrm{f}$) 
or number of frequencies ($\mathrm{N_{f}}$) are required as inputs to search periods 
in all unevenly spaced time series for all frequency finding methods. The 
$\mathrm{f_{min}}$ is usually defined as $2/T_{tot}$, where $T_{tot}$ is the
total timespan of the observations.
This definition is commonly used as a requirement to enclose at least two
variability cycles in the time series. However, variations having fewer than two
cycles can be considered with caution when biases have been identified and removed
\citep[e.g.][]{DeMedeiros-2013,FerreiraLopes-2015mgiant}. On the other hand,
$\mathrm{f_{max}}$ will be linked with the time interval between the
observations $\delta t$. The Nyquist frequency ($\mathrm{f_{max}} = 0.5/\delta
t$) must be assumed for evenly-spaced time series since this constitutes an
upper limit to the frequency range over which a periodogram can be uniquely 
calculated. Otherwise, for irregularly sampled cases, the frequency limit
becomes dominated by the exposure time \citep{Eyer-1999}.

The frequency sampling strategy is crucial 
to determine $\mathrm{f_{true}}$ using any frequency finding algorithm. A small
variation on $\mathrm{f_{true}}$ provides a big variation in the phase diagram
mainly when $f_{true}\times\,T_{tot} >> 1$ (see Sect.~\ref{sec_freqsamp}).
It means that $\mathrm{f_{true}}$ can be missed if the periodogram is not 
computed for a sufficiently large number of test frequencies. A reasonable
criterion (see previous paragraph) has been used to determine $\mathrm{f_{min}}$
while an empirical criterion has been applied to set $\mathrm{f_{max}}$
and $\Delta \mathrm{f}$. For instance, $\mathrm{f_{min}} = 0$,  $\mathrm{f_{max}} =
10d^{-1}$, and $\Delta \mathrm{f} = 0.1/T_{tot}$ were adopted by 
\citet[][]{Debosscher-2007} and \citet[][]{Richards-2012}. In this case, the 
authors assumed an empirical cutoff on the maximum frequency below
which any reliable frequency could be found: the frequency finding
method is able detect all reliable frequencies in a range of $\mathrm{f} \pm 
\Delta \mathrm{f}/2$.
On the other hand, \citet[][]{Schwarzenberg-Czerny-1996} assumes $\mathrm{f_{min}} = 0$, 
$\mathrm{f_{max}} = 1/2\tau_{med}$, and an optimal grid $\Delta \mathrm{f} = 
1/(A\times T_{tot})$, where $\tau_{med}$ is the median difference between 
successively ordered observation times and $A$ is a factor, typically ranging
$10-15$, which takes into account oversampling and binning or the number of harmonics
used in the Fourier fit. \citet[][]{Graham-2013} tested $\Delta \mathrm{f}$
values of $0.0001$, $0.001$, $0.01$, and the optimal grid over a frequency
range from $\mathrm{f_{min}} = 0$ to $\mathrm{f_{max}} = 20$ for standard 
frequency finding methods; LS, GLS, AOVs, PDMs, STR, FC, CE, SS, and CKP 
methods. The data test used by the authors has a number of observations ranging 
from $105$ to $966$ and a total baseline ranging from $2182$ to $2721$ days. 
The performance found for $\Delta \mathrm{f} = 0.0001$ and the optimal $\Delta
\mathrm{f}$ (the median optimal $\Delta \mathrm{f}$ is $2.5 \times 10^{-5}$) is
quite similar for all methods analysed according to the authors. Indeed
these results can only be used as a guide for samples that mimic those tested
by the authors since the samples tested do not cover all possible intervals of
measurements and baselines.
Therefore, what is the optimal resolution able to detect all periodic 
variations and how much finer grain is necessary to get a more accurate period
estimation, if $\mathrm{f_{true}}$ is found since an initial value can be found
with a coarser grain resolution, are open questions.

The majority of frequency 
finding methods were designed for single time series. Such methods are in
accordance with past surveys since they were usually from observations
in a single photometric waveband (e.g. VVV -
\citealt[][]{Minniti-2010}, LINEAR - \citealt[][]{Sesar-2011}, CoRoT - \citealt[][]{Baglin-2007}, and Kepler - 
\citealt[][]{Borucki-2010}). However, in the last few years
there are multi-wavelength surveys like Gaia \citep[][]{BailerJones-2013}, where
the main catalogue is multi-epoch using a wide G filter, but it also contains
colour information from simultaneous multi-epoch low resolution spectra. Period 
finding could be done on G-band and forthcoming surveys like LSST 
\citep[][]{Ivezic-2008} require multi-wavelength frequency finding methods  in
order to optimize the period searches. Usually each waveband is analysed 
separately and posteriorly the results are combined 
\citep[e.g.][]{Oluseyi-2012,FerreiraLopes-2015wfcam}. However, the combination
of different datasets allows us to increase the number of measurements that are
extremely important to signal detection.
\citet[][]{Suveges-2012} used principal component analysis to extract the best
period from analysis of the Welch-Stetson variability index 
\citep[][]{Welch-1993}. However, the method requires observations taken
simultaneously. \citet[][]{VanderPlas-2015} introduce a multi-band periodogram 
by extending the Lomb-Scargle approach. For that, the authors modeled each 
waveband as an arbitrary truncated Fourier series using the Tikhonov
regularization in order to provide a common model at all wavebands.
Such methods and new insights into multi-wavelength frequency finding methods 
are required to take full advantage of the multi-wavelengths observations.

The discussion above provokes questions that must be addressed in the challenge
to analyze large amount of photometric data automatically. Some of these
questions are addressed in the current paper (III) and the forthcoming papers
of this series will address the remaining questions. Sections
\ref{sec_freqsamp} and \ref{sec_freqerror} assess the frequency sampling 
and frequency uncertainties. Sect. \ref{sec_error} establishes an approach to
compute period and amplitude variations. In Sect. \ref{sec_results} we
show our results on estimating frequency resolution and uncertainties, and
discuss them. We address our final remarks in Sect. \ref{sec_conclusion}.

\section{Frequency sampling}\label{sec_freqsamp}

Consider a periodic signal modeled by function $\mathrm{F}$ having 
frequency $\mathrm{f_{true}}$ (being a real, positive constant) where 
$\mathrm{F = \left[F(\,t_{1}\,),F(\,t_{2}\,),\cdots,F(\,t_{n}\,)\right]}$. From
which $\mathrm{F(\,t_{i}\,) = F(\,t_{i} + n_{c}/\mathrm{f_{true}}\,)}$ where 
$n_{c}$ (number of cycles) is a positive integer ranging from
zero to $G[T_{tot}\times f_{true}]$. This relationship is also true
for phase values, i.e. $\mathrm{\phi_{i}(\,t_{i}\,) = \phi(\,t_{i} + 
n_{c}/\mathrm{f_{true}}\,) = \phi_{j}(\,t_{j}\,)}$ and therefore $\left|
\mathrm{\phi_{j}\,-\,\phi_{i}}\right| = 0$. The phase
difference between them for a frequency given by $\mathrm{f = f_{true} + \delta
f}$ is written as,

\begin{eqnarray}
 \scriptsize
 \left| \mathrm{\phi_{j}\,-\,\phi_{i}}\right| &=& \Bigl| \mathrm{t_{j}\times\left(f_{true}+\delta f\right) - G\bigl[ t_{j}\times \left(f_{true}+\delta f\right) \bigr]}   \nonumber \\
        & &  \mathrm{-t_{i}\times\left(f_{true}+\delta f\right) +G\left[ t_{i}\times \left(f_{true}+\delta f\right) \right]}  \Bigr| \nonumber \\
\left| \mathrm{\phi_{j}\,-\,\phi_{i}}\right| &=& \Bigl|\mathrm{\left(t_{j}-t_{i}\right)\times\delta f + \Bigl\{ t_{j}\times f_{true} - G \left[ t_{j}\times \left(f_{true}+\delta f\right) \right]  \Bigr\} }  \nonumber \\ 
     & & \mathrm{-\Bigl\{ t_{i}\times f_{true} - G \left[ t_{i}\times \left(f_{true}+\delta f\right) \right]  \Bigr\} \Bigr|}.          
 \label{eq_proof01}  
\end{eqnarray}
 
Having $\mathrm{\delta f << f}$ implies that $\mathrm{G \left[ t\times
\left(f+\delta f\right) \right] = G \left[ t\times f \right]}$. Indeed, this is
reasonable since the frequency sampling is usually set as $\mathrm{f_{n} =
f_{min}+n\times\delta f}$. For instance, $n=100$ implies that there is
a frequency at least hundred times greater than $\mathrm{\delta f}$.
Considering this limit, the two last terms (in curly brackets) of Eq.
\ref{eq_proof01} are cancelled and so,

\begin{equation}
 \mathrm{\bigr|\phi_{j}\,-\,\phi_{i}\bigr| \simeq \bigr|t_{j}-t_{i}\bigr|\times\delta f}   \implies  \mathrm{\delta f  \simeq \frac{\delta_{\phi_{j,i}}}{\bigr|t_{j}-t_{i}\bigr|}}
 \label{eq_deltaf}        
\end{equation}

The maximum variation on $\mathrm{\delta_\phi}$ is found for 
$\left|t_{j}-t_{i}\right| = T_{tot}$, i.e. from the comparison among the
measurements at the ends of the time series. Indeed, Eq. \ref{eq_deltaf} was
found only assuming that $\mathrm{\delta f << f}$ and hence this expression
can be used as an analytical definition of the frequency rate limit, where the
number of frequencies is given by:

\begin{eqnarray}
 N_{f} &=& \frac{f_{max}-f_{min} }{ \delta f } = \frac{\left(f_{max}-f_{min}\right)\times T_{tot}}{\delta_\phi } \nonumber \\
   N_{f}  &\simeq& \frac{f_{max}\times T_{tot}}{\delta_\phi},
 \label{eq_nfmax}        
\end{eqnarray}

\noindent where $\mathrm{f_{max} >> f_{min}}$ was assumed when deriving the
expression. This expression enables us to determine $N_{f}$ from the phase
shift $\delta_\phi$ since $T_{tot}$ is a feature of a time series. On the other
hand, $\mathrm{f_{max}}$ can be assumed to be the same for different
time series, in the same set of observations,
since the upper limit of frequencies, for those time series
having evenly spaced data, is given by the Nyquist frequency
\citep[e.g.][]{Eyer-1999}.
Therefore, the frequency search will be performed with the same resolution in 
the phase diagram if we assume the same $\delta_\phi$ for different time series.
Moreover, it facilitates a strict comparison of frequency finding searches
performed by different surveys.

Equation \ref{eq_nfmax} was defined only by considering the phase
diagram. Therefore, this relation is general and it can be used as an
accurate determination of the frequency grid required to find any signal.
Indeed, a similar equation has been used to estimate the frequency grid 
\citep[e.g.][]{Schwarzenberg-Czerny-1996,Debosscher-2007,Richards-2012,
VanderPlas-2015,VanderPlas-2018} where the $1/\delta_\phi$ is called the
oversampling factor. However, no meaning had been provided for the
oversampling factor so far. Values ranging from 5 to 15 have being  adopted
empirically only to ensure that the frequency grid is sufficient to sample each
periodogram peak. The proper meaning of the oversampling factor is
defined in Eq.
\ref{eq_nfmax} from which a suitable frequency grid for any kind of signal can
be determined.

\begin{figure}
  \centering
  \includegraphics[width=0.49\textwidth,height=0.4\textwidth]{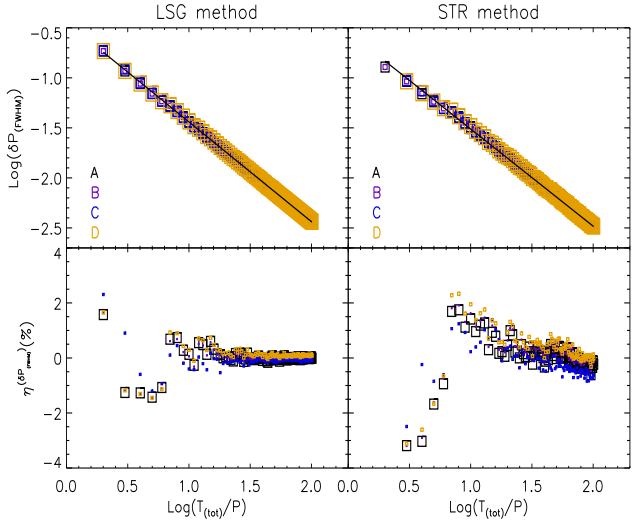} 
  \caption{The logarithmic of FWHM as a function of logarithmic of 
  $T_{tot}/P$ for the Generalized Lomb-Scargle (LSG -  upper left
  panel) and for the string length (STR upper right panel)
  methods using four (ABCD) sinusoidal signal having models of
  variation and noise, shown in Fig~\ref{fig_discerror} and described in  Sect.
  \ref{sec_error}.
  The models are set by colors and the solid black line marks the linear fits for LSG and STR 
  methods. The percent relative errors ($\eta$) for both results are shown in
  the lower panels.}
  \label{fig_fhwm}
\end{figure}

\section{Frequency uncertainties}\label{sec_freqerror}

Frequency uncertainties were analytically defined from Fourier analysis 
\citep[e.g.][]{Kovacs-1981,Gregory-2001,Stecchini-2017},

\begin{equation}
 \sigma_{f} \propto \frac{1}{T_{tot}}\sqrt{\frac{1}{n\times\Sigma}}
 \label{eq_olderror}        
\end{equation}

\noindent where $n$ is the number of measurements, $\Sigma$ is the 
signal-to-noise ratio, and $T_{tot}$ is the total baseline of the observations.
The uncertainty provided by a well defined periodic signal will be limited
by the exposure time and hence Eq. \ref{eq_olderror} is not a suitable
definition since it assumes an infinite accuracy.
On the other hand, phenomena which result in small variations on the period can
be mistaken for an increased uncertainty. Indeed, the uncertainties computed
using a time series are given by the sum of intrinsic plus instrumental
limitations. The uncertainties related with instrumental limitations can be
estimated using a noise model \citep[e.g.][]{FerreiraLopes-2017papII} and by
including this we can thus estimate the intrinsic variation.
Some inconsistencies are found when the frequency
uncertainty ($\sigma_{f}$  - see Eq.~\ref{eq_olderror}) estimation is related
with $T_{tot}$, $n$, and $\Sigma$.
For instance, a signal having an intrinsic variation in the frequency 
($\sigma_{f} \ne 0$), such as light curves of rotational variables, may return a
similar estimation of the uncertainty for time series having one hundred or one
thousand measurements.
On the other hand, for periodic signals a reduction in the dispersion about the 
model naturally occurs for a longer baseline and its accuracy is limited by 
instrumental properties instead for large $T_{tot}$ or $\Sigma$.
Indeed, the power spectrum of $\mathrm{f_{true}}$ 
tends to a delta function with increasing $T_{tot}$ while increasing $n$ and 
$\Sigma$ improves the signal reliability since the probability of a signal being
detected increases when the noise is reduced. These properties
characterize the signal but they are not directly related with the period variations.

The \textit{CoRoT} and \textit{Kepler} databases have in common a large
number of measurements and wide coverage time that provide unreliable uncertainty 
estimations using Eq. \ref{eq_olderror}. Therefore, new approaches have been
used to compute frequency uncertainties for semi-periodic signals. For
the \textit{Kepler} time series, \citet[][]{Reinhold-2013} compute the
frequency uncertainty by fitting a parabola to the peak of the Lomb-Scargle power spectrum
(Reinhold, private communication). On the other hand, 
for the \textit{CoRoT} time series, \citet[][]{DeMedeiros-2013} used a
similar equation to that proposed by \citet[][]{Lamm-2004} to estimate the period
uncertainty, given by

\begin{equation}
 \delta P = \frac{\delta\nu \times P^2}{2},
 \label{eq_eperlamm}  \nonumber      
\end{equation}

\noindent where $\delta\nu$ is about $1/T_{tot}$ for 
non-uniform sampling according to the authors.
\citet[][]{FerreiraLopes-2015mgiant, FerreiraLopes-2015cycles} also used
the \textit{CoRoT} time series to study non-radial pulsation and stellar
activity where the period uncertainties were estimated as the
\textit{FWHM} ($\delta P_{(\rm FWHM)}^{(\rm STR)}$) of the String Length power
spectrum \citep[][]{Dworetsky-1983}. In particular, the amplitudes and
periods vary for light curves of rotational variables that have differential
rotation and spot evolution \citep[e.g.][]{Lanza-2014, Reinhold-2015,DasChagas-2016}. The analytical 
expression given by Eq.~\ref{eq_olderror} or the analysis of the power spectrum 
are half-way to computing period variations in order to get new
clues about physical phenomena that account for such variations.
 
Figure \ref{fig_fhwm} shows $\delta P_{(\rm FWHM)}$ as a function of
the number of cycles ($N_{(cycles)} = T_{tot}/P$) for the Generalized Lomb-Scargle 
\citep[LSG-][]{Zechmeister-2009} and for the string
length \citep[STR -][]{Dworetsky-1983} methods using the sinusoidal signal 
described in Sect. \ref{sec_error}. The best fit models found for the LSG and
STR methods are given by,

\begin{equation}
 Log\left(\delta P_{(\rm FWHM)}^{(\rm LSG)}\right) = -1.00 -0.44 \times Log\left(\frac{T_{tot}}{P}\right)
 \label{eq_eqfwhm_lsg}        
\end{equation}

\noindent and

\begin{equation}
 Log\left(\delta P_{(\rm FWHM)}^{(\rm STR)}\right) = -0.97 -0.54 \times Log\left(\frac{T_{tot}}{P}\right).
 \label{eq_eqfwhm_str}        
\end{equation}

\begin{figure}
  \centering
  \includegraphics[width=0.49\textwidth,height=0.9\textwidth]{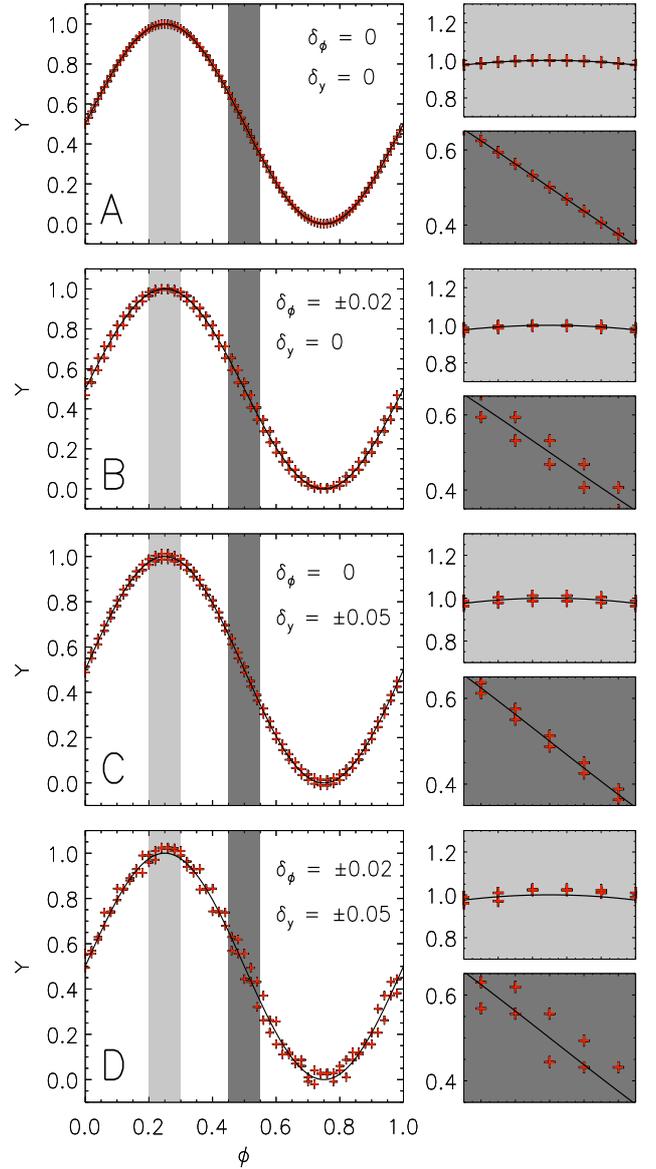} 
  \caption{Sinusoidal light curves, with a fixed period
  and amplitude (A); with a varying period ($\delta_{\phi} = 0.02$) and constant amplitude
   (B); varying amplitude ($\delta_{y} = 0.05$) and constant period (C);
   and both amplitude and period variations (D). The black line shows the model
   while the red crosses show the measurements. The dark and light grey regions 
   are expanded in the right panels.}
  \label{fig_discerror}
\end{figure}

We create 4 different sinusoidal models which are a
single sinusoid (A), sinusoid plus amplitude variation (B), sinusoid plus period
variation (C), and a sinusoid plus amplitude and period variations (D), see
Sect. \ref{sec_error} and Fig.~\ref{fig_discerror} for more
details.
However, the results are quite similar for all ABCD models, i.e. the 
$\delta P_{(\rm FWHM)}$ estimation is mainly defined by the $N_{(cycles)}$
instead of the time series properties.
Indeed, it is highlighted for  $N_{(cycles)} > 10$ where the
percent relative error (i.e. $\eta = 100\times(theoretical-experimental)/\mid 
theoretical \mid$) is always smaller than $4\%$, see lower panels of
Fig.~\ref{fig_fhwm}. Eq.
\ref{eq_eqfwhm_lsg} is slightly different from Eq. \ref{eq_eqfwhm_str} (see the
solid lines in the two upper panels of Fig. \ref{fig_fhwm}) but the LSG method 
shows smaller relative errors ($\eta$). The approach using the FWHM of the 
power spectrum and any frequency finding method does not provide a trustworthy 
estimation of the period variation (for more detail see Sect.
\ref{sec_testreal}).

To summarise, the uncertainty computed using the power
spectrum gives us a period interval about the variability period that leads to
similar phase diagrams. Indeed, the main period and its uncertainty can vary for different period 
finding methods since the susceptibility to measuring phase diagram variations
is not the same \citep[e.g.][]{Eyer-2006,Graham-2013}. 
Moreover, the main period is assumed to be one that leads to the highest peak of
periodogram that, a priori, gives the smoothest phase diagram and also the
smallest residuals ($\sigma_r$), i.e. the standard deviation of observed data
minus model (or predicted value). However, these assumptions have not been
analysed so far, but this empirical criteria has been used all the time. In the
sections below these questions are addressed.

\section{Frequency and Amplitude Variations}\label{sec_error}

\begin{figure*}
  \centering
  \includegraphics[width=0.9\textwidth,height=0.9\textwidth]{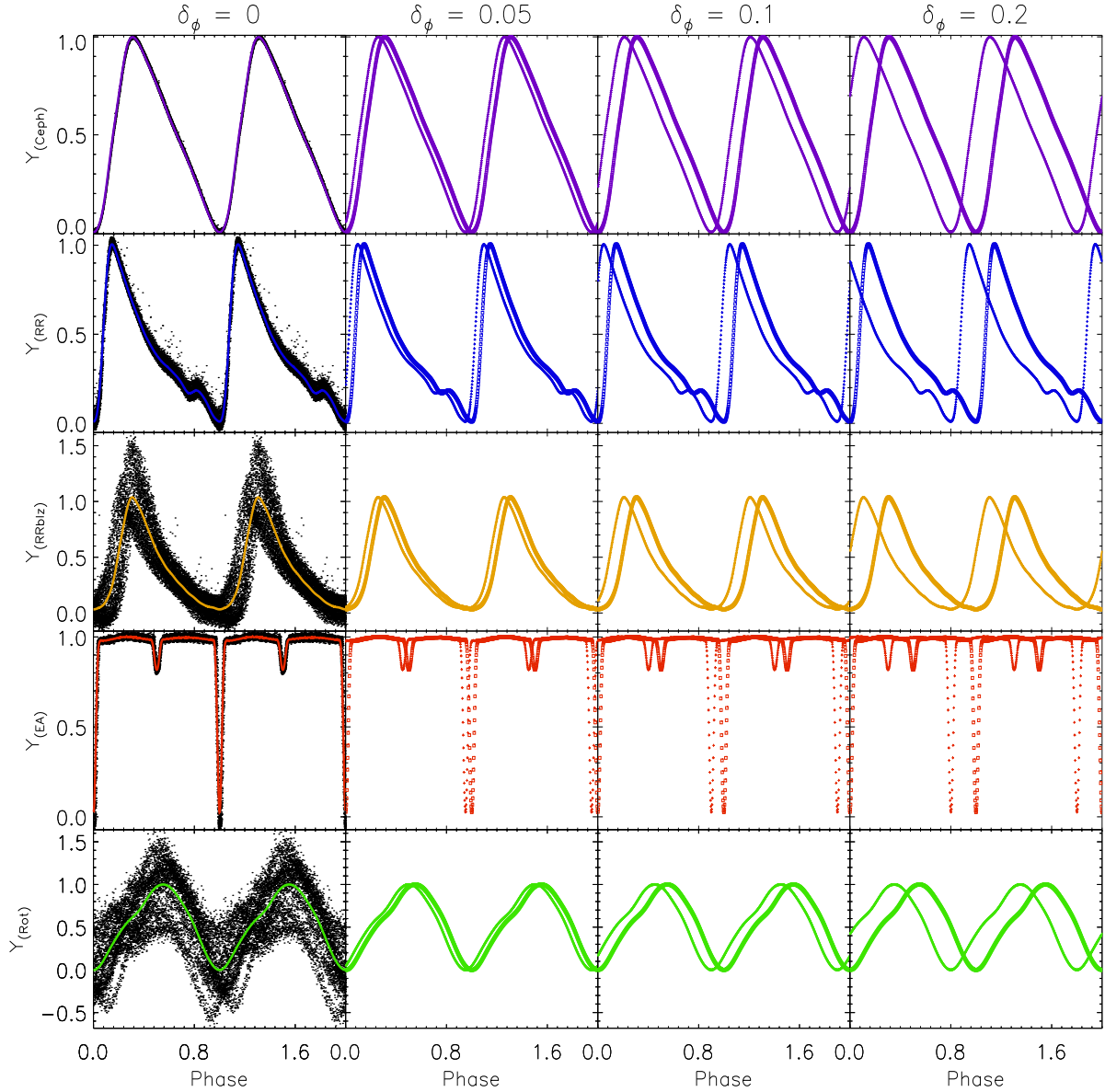} 
  \caption{Ceph (purple - \textit{CoRoT-211626074}), RR (blue - \textit{CoRoT-101370131}),
   RRblz (yellow - \textit{CoRoT-100689962}), EA (red -
   \textit{CoRoT-102738809}), and Rot (green - \textit{CoRoT-110843734}) phase diagrams.
  The dark dots in the first panels indicate the original data while the solid
  lines set the Fourier models $M(\phi)$.
  The next panels were build from the models
  where half of them are set by  squares (measurements at $t > t_{0}+1./f$) 
  and another half by crosses (measurements at $t > T_{tot}-1./f$).  
  The frequencies used to build the phase diagram from left to
  right panels are $\mathrm{f_{true} + \delta f}$ to  $\delta_\phi = 
  [0,0.05,0.1,0.2]$, respectively. }
  \label{fig_lctype}
\end{figure*}

Eq.~\ref{eq_olderror} is an analytical expression that saves computational time.
However, computational methods can be used to perform non-analytical approaches
to compute frequency and amplitude variations in order to accurately 
choose the main variability period as well as give additional information about
the phenomena observed. Indeed, any time series can be modeled using Fourier 
decomposition ($Y(\phi)$ - see Fig. \ref{fig_lctype}). In order to determine the
suitable measurements or light curve regions to compute these
variations consider the following example:
 
\begin{equation}
    t^{(o)} = t^{(m)} + \delta_{t} \qquad $and$ \qquad y^{(o)} = \sin\left(2 \times \pi \times t^{(m)} \times \mathrm{f_{true}} \right)+\delta_{y}
 \label{eq_linmodel}    
\end{equation} 

\noindent where $(o)$ and $(m)$ are the mean observed and modeled values,
respectively. Indeed, $t^{(o)} = t^{(m)}$ if $\delta_{\phi} = 0$ and $y^{(o)} =
y^{(m)}$ if $\delta_{y} = 0$. Four cases are displayed in the Fig.
\ref{fig_discerror}; (A) constant period and amplitude, (B) period
variation for constant amplitude, (C) amplitude variation for constant period, 
and (D) period and amplitude variation. It is easier to understand these cases
if a linear fit is calculated in the light and dark grey regions in Fig.
\ref{fig_discerror}, given by

\begin{equation}
    y^{(o)} =  \alpha \times \left( \phi + \delta_{\phi}^{(e)} \right) + \left( \beta + \delta_{y}^{(e)} \right)
 \label{eq_line}    
\end{equation} 

\noindent where $(e)$ means expected value and ($\alpha,\beta$) becomes 
($\alpha_{y},\beta_{y}$) or ($\alpha_{\phi},\beta_{\phi}$) to indicate the 
region used to estimate the amplitude or period variations, respectively.
Moreover, $y^{(o)} = y^{(m)}$ if $\delta_{\phi}^{(e)} = 0$ and 
$\delta_{y}^{(e)} = 0$. The linear fit only takes into account the
first order contribution. However, this allows us to determine
a simple analytical equation to analyse the contributions of
$\delta_{\phi}^{(e)}$ on $\delta_{y}^{(e)}$.
For real data, the fitting function, that may have a more complex shape, can be
used. The main features derived from Fig. \ref{fig_discerror} are summarized
below:

\begin{itemize}

 \item[\textit{i} -- ] The computed amplitude variation ($\delta_{y}^{(c)}$) is
 defined as the difference between the observed ($o$) and modeled ($m$) 
 amplitude at the same phase, i.e. $[\phi^{(o)},y^{(o)}]$ implied from $y^{(m)}
 = Y(\phi^{(o)})$, is given by

  \begin{equation}
     \delta_{y}^{(c)} = y^{(o)}-y^{(m)} = \alpha_{y} \times \delta_{\phi}^{(e)} + \delta_{y}^{(e)}.
   \label{eq_uncamp}    
  \end{equation} 

$\delta_{y}^{(c)}$ can be different from zero if $\delta_{y}^{(e)}
= 0$ according to Eq. \ref{eq_uncamp}, i.e. a phase variation $\delta_{\phi}^{(e)}$ 
can appear as an amplitude
variation if $\alpha_{y} \ne 0$. On the other hand, the ideal case will be found
when $\delta_{y}^{(c)} = \delta_{y}^{(e)}$ that implies that $\alpha_{y} \times
\delta_{\phi}^{(e)} \simeq 0$. For the cases where  $\delta_{y}^{(e)} \ne 0$ the
ratio of computed to expected values is written as
  
  \begin{equation}
    R^{(\delta_{y})} = \left| \frac{\delta_{y}^{(c)}}{\delta_{y}^{(e)}} \right| = \left|\alpha_{y} \times \frac{\delta_{\phi}^{(e)}}{\delta_{y}^{(e)}} + 1\right|
   \label{eq_uncamprel}    
  \end{equation} 
  
\noindent that indicates whether the computed value is overestimated ($R^{(\delta_{y})} >
1$), equal ($R^{(\delta_{y})} = 1$), or underestimated ($R^{(\delta_{y})} < 1$).
Therefore, the estimation of $\delta_{y}^{(e)}$ will be improved if the 
$\alpha_{y} \times \delta_{\phi}^{(e)} << \delta_{y}^{(e)}$.
$\delta_{\phi}^{(e)}$ is a time series property that can not be modified. However
those light curve regions having $\alpha_{y} \simeq 0$ provide a better estimation
of the amplitude variation. The
light-grey region of Fig. \ref{fig_discerror} indicates a suitable region to
measure amplitude variation since this contains the smallest $\alpha_{y}$
values.
 
 \item[\textit{ii} -- ] The computed phase variation ($\delta_{\phi}^{(c)}$) is given
 by the difference between the observed and modeled value for the same
 amplitude, i.e. for each pair of observed measurements $[\phi^{(o)},y^{(o)}]$ 
 implies that $y^{(o)} = Y(\phi^{(m)})$, which can be written as

  \begin{equation}
     \delta_{\phi}^{(c)} = \phi^{(o)}-\phi^{(m)} = \frac{\delta_{y}^{(e)}}{\alpha_{\phi}} + \delta_{\phi}^{(e)}.
   \label{eq_unper}    
  \end{equation} 
 
\noindent where $\delta_{\phi}^{(c)}$ will return values different to zero
even if $\delta_{\phi}^{(e)}=0$ in the same fashion as the amplitude variation
(see item \textit{i}). Indeed, the amplitude and phase variations are coupled 
equations, i.e. $\delta_{y}^{(c)}$ depends on $\delta_{\phi}^{(e)}$ while
$\delta_{\phi}^{(c)}$ depends on $\delta_{y}^{(e)}$ (see Eqs. \ref{eq_uncamp}
and \ref{eq_unper}).  Moreover, not all observed measurements can be used to 
compute $\delta_{\phi}$ since those having values bigger or smaller than the
maximum and minimum $Y$ values cannot be written as $y^{(o)} = Y(\phi^{(e)})$.
Therefore, only the observed measurements having values between the minimum ($Y_{min}$)
and maximum ($Y_{max}$) model values can be used to estimate $\delta_{\phi}^{(c)}$,
i.e. for all $y^{(o)}$ since $Y_{min} < y^{(o)} < Y_{max}$. The number of
measurements  used to compute $\delta_{\phi}^{(c)}$ will depend on the signal
type (see Fig. \ref{fig_lctype} first panels). However these measurements do not
provide a good information of time variation about the model. Using the ratio of
computed and expected values is a suitable way to examine agreement between
them, given by 
  
  \begin{equation}
    R^{(\delta_{\phi})} =  \left| \frac{\delta_{\phi}^{(c)}}{\delta_{\phi}^{(e)}} \right| = \left| \alpha_{\phi}^{-1} \times \left( \frac{\delta_{\phi}^{(e)}}{\delta_{y}^{(e)}} \right)^{-1} + 1 \right|.
   \label{eq_unperrel}    
  \end{equation}

  The opposite result of $R^{(\delta_{y})}$ is found since the
  dispersion of $\delta_{\phi}$ values are proportional to the inverse of
  the angular coefficient and to the inverse of the relation
  $\delta_{\phi}^{(e)}/\delta_{y}^{(e)}$. This means that the weight of
  $\delta_{y}^{(e)}$ on $\delta_{\phi}^{(c)}$ and $\delta_{\phi}^{(e)}$ on 
  $\delta_{y}^{(c)}$ will be the same only if
  $\delta_{\phi}^{(e)}/\delta_{y}^{(e)} = 1$. The regions of the light curve
  where the highest $\alpha_{\phi}$ values are found will be better to compute
  $\delta_{\phi}^{(c)}$ since the weight of
  $\delta_{\phi}^{(e)}/\delta_{y}^{(e)}$ is minimized. For instance, the 
  highest $\alpha_{\phi}$ values for the sinusoidal variation will be found in
  the dark-grey region of  Fig. \ref{fig_discerror}. 
  
 \item[\textit{iii} -- ] Fig. \ref{fig_discerror} B shows a sinusoidal
 light curve considering $\delta_{\phi}^{(e)} = 0.02$ and $\delta_{y}^{(e)} =
 0$. As expected $\delta_{y}^{(c)} \simeq 0$ is in the flattest region of
 the light curve. A note of caution, these regions are not well modeled by a
 straight line and non-linear effects, different from those analysed in items \textit{i}
 and \textit{ii}, can be found. Therefore, a fit to the whole light curve
 rather than a linear fit is necessary. The best estimation of the
 amplitude variation will be found if the region is small enough so that the
 model and linear fit are in agreement.
 Indeed, the periodic variation region can be approximately described by 
 a linear model but the estimation of $\delta_{\phi}^{(c)}$
 is computed using the time series model (see Sect. \ref{sec_visualdeltaphi}). 
 The size and complexity of regions used to measure the period and amplitude variations
 are strongly dependent on the time series signal.
 To summarize, there is a non-zero contribution to the phase variation of the
 estimation of amplitude variations, if the region cannot be modeled by a
 horizontal line.
 On the other hand, $\delta_{\phi}^{(c)}$ is accurately estimated from  Eq.
 \ref{eq_unper} since for this example $\delta_{\phi}^{(c)} = \delta_{\phi}^{(e)}$.

 \item[\textit{iv} -- ] A sinusoidal light curve considering
 $\delta_{\phi}^{(e)} = 0$ and $\delta_{y}^{(e)} = 0.05$ is shown in Fig. 
 \ref{fig_discerror} C. From the approach described in item \textit{i} the
 estimation of $\delta_{y}^{(c)}$ is accurately estimated from Eq.
 \ref{eq_uncamp} since for this example $\delta_{y}^{(c)} = \delta_{y}^{(e)}$.
 On the other hand, $\delta_{\phi} \ne 0$ despite $\delta_{\phi}^{(e)} = 0$ for
 the the dark grey region in  Fig. \ref{fig_discerror}. Indeed,
 $\delta_{\phi}$ will be equal to  $\delta_{\phi}^{(e)}$ for the case where 
 $\delta_{y}^{(e)} \ne 0$ only when $\alpha_{\phi} = \infty$, i.e a
 perpendicular line to the phase axis. Indeed, the phase variation is
 dominated by the amplitude variation in these cases since
 $\delta_{y}^{(e)}/\alpha_{\phi} >> \delta_{\phi}^{(e)}$.

\item[\textit{v} -- ] Fig. \ref{fig_discerror} D shows the sinusoidal light
curve when $\delta_{\phi}^{(e)} = 0.02$ and $\delta_{y}^{(e)} = 0.05$. It
exemplifies a real time series where some variation in time and flux is
always found. However the  ratio of $\delta_{\phi}^{(e)}/\delta_{y}^{(e)}$ 
will determine the relative weights of each other (see Eqs.
\ref{eq_uncamprel} and \ref{eq_unperrel}).
For the current example $R^{(\delta_{y})} = \left| 1 + 0.4 \times \alpha_{y} \right|$  and
$R^{(\delta_{\phi})} =  \left| 1 + 2.5 \times \alpha_{\phi}^{-1} \right|$. Therefore, the
best scenario to compute period and amplitude variations is the one
where $\alpha_{y} = 0$ and $\alpha_{\phi} = \infty$. However, this is usually
not the case, and hence such variations will not be computed precisely. Therefore,
Monte-Carlo simulations are performed in Sect. \ref{sec_results} in order to estimate the inaccuracy of
$\delta_{y}^{(c)}$ and $\delta_{\phi}^{(c)}$ as proxies for the variation
on $\delta_{y}^{(e)}$ and $\delta_{\phi}^{(e)}$.

\end{itemize}

The discussion above does not take into account any particular light curve shape
and hence this argument can be applied to all light curves types. Moreover, multiple
regions of the phase diagram can be used to compute the amplitude and
period variations.
Indeed, these regions must be chosen following the discussion above in order to
minimize the amplitude on period variations and vice versa, i.e use the flattest
regions to compute the amplitude variation and the regions with the largest
gradients (positive or negative) for the period variations.
Indeed, time series having saddle regions also are suitable to 
compute the amplitude variation for the same reason discussed above (see panel A in Fig. \ref{fig_lcreg}). 
A more detailed description about how to compute the $\delta_{y}^{(c)}$ and $\delta_{\phi}^{(c)}$ is presented in 
Sect. \ref{sec_computeerror}.

\subsection{Computing period and amplitude variations}\label{sec_computeerror}

Consider a generic light curve modeled by $Y\left(\phi\right)$ for $\left[
\phi_{1},\phi_{2},\cdots,\phi_{N} \right]$ where $\phi_{i}$ are in ascending
order. The tangent angles to the model can be written as

  \begin{equation}
     \theta_{i} =  \arctan \left(\frac{   Y\left( \phi_{i+1} \right) - Y\left( \phi_{i} \right) }{ \phi_{i+1} - \phi_{i}} \right).
     \label{eq_atan}    
  \end{equation}

The angles are better to use than the $\alpha$ values to determine suitable
regions to compute the period and amplitude variations because they can
be assessed from the model without making any additional computation (see Sect.
\ref{sec_error}). The largest $\theta_{i}$ values are associated with the largest $\alpha$ values and
the smallest $\theta_{i}$ values are associated with the smallest $\alpha$
values. The regions having smaller or bigger angles will be better to compute
$\delta_{y}$ and $\delta_{\phi}$, respectively (see Sect. \ref{sec_error}). 
After defining the region to compute these variations the following
procedures should be performed:

\begin{itemize}

   \item Calculate the amplitude variation ($\sigma_{y}$): consider the
   region(s) that enclose the majority of measurements having
   $\mathrm{|\theta| < \theta_y}$. Next for each 
   $[\phi_{i}^{(o)},y_{i}^{(o)}]$ we find its respective 
   $[\phi_{i}^{(m)},y_{i}^{(m)}]$ from which the vector $\delta_{y} = \left[ y_{i}^{(o)}-y_{i}^{(m)}
   ,\cdots,y_{N}^{(o)}-y_{N`}^{(m)}\right]$ is obtained.
   Lastly, the amplitude variation is computed as

\begin{equation}
  \sigma_{y} = \gamma \times eMAD\left( \delta_{y_{i}} \right)
 \label{eq_neweamp}        
\end{equation} 

\noindent where $\rm eMAD$ is the even-median absolute 
deviation of the even-median \citep[][]{FerreiraLopes-2017papII} and $\gamma$ is
a correction factor (for more detail see Sect. \ref{sec_sortreg}).
The $\rm eMAD$ is a slight modification to the $\rm MAD$ (the median absolute
deviation of median). Indeed, $\sigma_{y}$ becomes a robust estimate of the
standard deviation to outliers if $\gamma =  1.48$ according to \citet[][]{Hoaglin-1983}.
A note of caution, $\delta_{y}$ is computed using $y_{i}^{(m)}$ instead of
$y_{i}^{(e)}$ since the first one provides better estimations of expected
values if the region cannot be well modeled by a line. Indeed,
$y_{i}^{(m)} \simeq y_{i}^{(e)}$ only if $\theta_{i} \approx 0$.
   
  \item Calculate the period variation ($\sigma_{P}$): consider the region
  that encloses the majority of measurements having $\mathrm{|\theta| >
  \theta_P}$. For each $[\phi_{i}^{(o)},y_{i}^{(o)}]$ we find its respective
  $[\phi_{i}^{(e)},y_{i}^{(e)}]$, 
  from which the vector $\delta_\phi = \left[ \phi_{i}^{(o)}-\phi_{i}^{(e)} ,\cdots, \phi_{N}^{(o)}-\phi_{N}^{(e)}\right]$ 
  is obtained. Lastly, the period variation is computed as

  \begin{equation}
     \sigma_{P} = \gamma \times P \times eMAD\left( \delta_{\phi_i} \right).
     \label{eq_neweperr}        
   \end{equation}

\end{itemize}

The current approach estimates the period and amplitude uncertainties taking
into account the variations about a model. Eqs. \ref{eq_neweamp} and
\ref{eq_neweperr} are computed using only those measurements suitable to reduce
the weight of either $\delta_{y}$ or $\delta_{\phi}$. However, the
accuracy of $\sigma_{y}$ and $\sigma_{P}$ are extremely dependent on $\theta_y$
and $\theta_P$, respectively. For instance, values of $\theta = 
[0.1^\circ,1^\circ,5^\circ,10^\circ,70^\circ,80^\circ,89^\circ,89.9^\circ]$ 
return $\alpha^{(\phi)} =
[\sim0.002,\sim0.02,\sim0.09,\sim0.2,\sim2.8,\sim5.7,\sim57,\sim573]$. Indeed, 
the optimal choice of $\theta$ values is a compromise between the number of 
measurements enclosed for each limit and the usefulness of these
measurements.
Moreover, the statistical significance increases with the number of measurements
while a higher signal-to-noise reduces the weight of $\delta_{y}$ on
$\delta_{\phi}$. Therefore, the number of measurements and 
signal-to-noise are indirectly implicated in the period and amplitude
variations.

\section{Results and Discussion}\label{sec_results}

Setting correct inputs using either method to select variable stars or to
perform frequency finding searches is mandatory to get accurate outputs. The
variability indices used to select variable stars candidates were studied 
deeply in the first two papers of this series,
\citet[][]{FerreiraLopes-2016papI, FerreiraLopes-2017papII}. These studies
enabled us to provide the optimal constraints on noise models and
establish well-defined criteria to settle the best approach to discriminate variable 
stars from noise as well as to affirm that the selection of a reliable sample 
is unfeasible using variability indices. Therefore, frequency analysis may also
be used to select out untrustworthy variations but all constraints must be
properly delimited and understood to avoid mistakes. For instance, the
interquartile range can provide an incorrect list of variable star candidates
if the time sampling is not taken into account. Therefore, all the
relevant points about frequency finding methods were discussed in
Sect.~\ref{introduction}. The $\mathrm{f_{min}}$ and $\mathrm{f_{max}}$ are limited by the time 
series and the maximum reliable frequency, respectively. On the other hand, the
sampling frequency was addressed in Sect. \ref{sec_freqsamp} in order to 
facilitate making a decision about the frequency resolution taking into account
the effects on the frequency search. The frequency sampling and a new approach
to computing period and amplitude variations are outlined in sections below.

\begin{table}
  \begin{threeparttable}
 \centering     
 \caption[]{Constraints on the frequency search analysis performed by different
 surveys. The designation, $\mathrm{f_{min}}$, $\mathrm{f_{max}}$, mean total 
 time span $\mathrm{\overline{T_{tot}}}$, and $\mathrm{N_{f}}$ of each survey 
 are shown. The frequency unit is day$^{-1}(d^{-1})$ and
 $\mathrm{\overline{T_{tot}}}$ is in days (d). Moreover,
 $\delta_{\phi}$ Eq.\ref{eq_deltaf} is given in the last column. 
 }\label{tabconstraint} \begin{tabular}{c c c c c c}        
   \hline\hline                 
    Survey & $\mathrm{f_{min}}(d^{-1})$ & $\mathrm{f_{max}}(d^{-1})$ & $\mathrm{\overline{T_{tot}}}(d)$ & $\mathrm{N_{f}}$ & $\delta_{\phi}$ \\   
   \hline     
    CoRoT & $2/T_{tot}$ & $3$  & $\sim 136$ & $2\times 10^{3}$ &  $0.20 $  \\ 
    GAIA & $2/T_{tot}$ & $3.9$  & $\sim 1700$ & $\sim3\times 10^{3}$ &  $0.19$  \\
    Kepler & $\sim 3/T_{tot}$ & $1$  & $\sim 90$ & $1300$ &  $0.07$  \\        
    OGLE & $0$ & $24$  & $\sim 2780$ & $10^{4}$ &  $ > 1$  \\ 
    TAROT$^{1}$ & $2/T_{tot}$ & $f_{max}$  & $\sim 900$ & $10^{5}$ &  $\sim 0.22$  \\ 
    WFCAM$^{2}$ & $2/T_{tot}$ & $f_{max}$  & $\sim 1058$ & $10^{5}$ &  $\sim
    0.25$  \\
   \hline                                   
 \end{tabular}
  \begin{tablenotes}
      \small
    \item 1 - The frequency step is taken as described in \citet[][]{Akerlof-1994} and \citet[][]{Larsson-1996}.
    \item 2 - $f_{max}$ computed according to \citet[][]{Eyer-1999}.
    \end{tablenotes}
  \end{threeparttable}
\end{table}

\subsection{Optimal frequency sampling}\label{sec_resolution}

An optimal determination of $\mathrm{f_{max}}$ is critical to reducing running
time since it leads to the determination of the resolution and thus the number
of frequencies or loops performed by the frequency finding algorithm (see
Eq.~\ref{eq_nfmax}). Estimation of $\mathrm{f_{max}}$ using the Nyquist
frequency for oversampled data returns an overestimated frequency,
i.e. frequencies that are this high are not reliably measured using the
available data. Indeed, for unevenly and poorly sampled time series, the
Nyquist frequency can be under or over-estimated whatever the estimation of the
time interval from the measurements (as a mean or median value). 
For instance, long and short cadence CoRoT light curves have
$\mathrm{f_{max}}$ of about $169d^{-1}$ and $2790d^{-1}$, respectively. These 
frequencies imply that the search for periodic variations at higher frequencies
will not be productive. Therefore, empirical values have been adopted as
the frequency limit. $\mathrm{f_{max}} = 10d^{-1}$ has been generally
adopted \citep[e.g.][]{Debosscher-2007,Richards-2012,DeMedeiros-2013} but higher values
also can be found \citep[e.g.][]{Schwarzenberg-Czerny-1996,Damerdji-2007,FerreiraLopes-2015wfcam}. 
The parameters used to perform frequency searches for
variable star catalogs for some surveys are listed in Table
\ref{tabconstraint}; the WFCAM multi-wavelength variable star catalog 
\citep[WFCAM - ][]{FerreiraLopes-2015wfcam}, the Optical Gravitational Lensing
Experiment \citep[OGLE -][]{Soszynski-2009}, the TAROT suspected variable star catalog 
\citep[TAROT -][]{Damerdji-2007}, 
GAIA\footnote[1]{https://gaia.esac.esa.int/documentation/GDR1} data release 1 documentation, the
semi-sinusoidal variables in the CoRoT mission \citep[SR-CoRoT
-][]{DeMedeiros-2013}, rotation periods of 12 000 main-sequence Kepler stars
\citep[Kepler -][]{Nielsen-2013}, and the WFCAM multiwavelength Variable Star 
Catalog \citep[WFCAM -][]{FerreiraLopes-2015wfcam}. The $\mathrm{f_{max}}$
adopted by OGLE was used to estimate $\delta_{\phi}$ for the WFCAM and TAROT
catalogs. Indeed, $\mathrm{f_{max}}$ values given by analytical expressions in
\citet[][]{Eyer-1999}  depend on each time series and such values are usually 
much higher than those empirically adopted.

The frequency sampling defined by Eq. \ref{eq_nfmax} was designed without taking
into account any particular criteria and hence this expression may work for any
signal type. Indeed, the number of constraints is not reduced, but 
the frequency sampling given by shifts on the phase $\delta_{\phi}$ instead
of shifts in frequency is clearer to read. Moreover, Eq. \ref{eq_deltaf} also
enables us to determine how much finer grain resolution is required to get a more accurate
frequency estimation if the variability frequency is found since an initial
value can be found with a coarser grain resolution. The frequencies not included
in the frequency sampling may be detected or not, depending on the response to
the frequency finding method for frequencies given by $\mathrm{f \pm \delta f/2}$,
for instance. Indeed, the resolution of frequency sampling is critical for
a large $T_{tot}$ since we find larger variations in the phase diagram 
for nearby frequencies. Moreover, as highlighted in previous sections, 
$\delta_{\phi}$ standardizes the criteria to perform frequency searches for 
time series having different total time spans. It allows us to compare 
straightforwardly the frequency analysis performed in different photometric surveys.

 \begin{figure*}
  \centering
  \includegraphics[width=0.9\textwidth,height=0.9\textwidth]{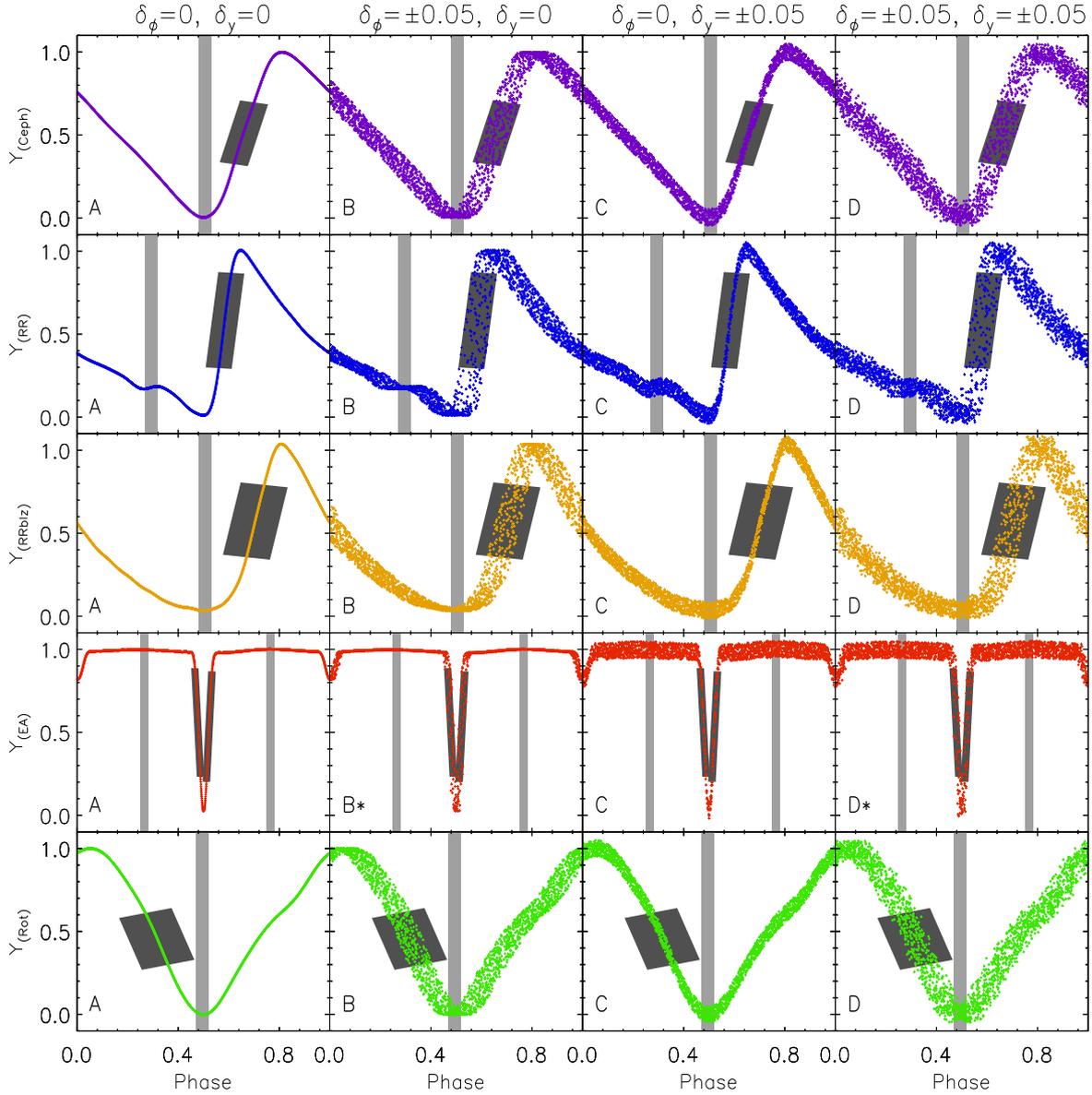} 
  \caption{A single period (in contrast to Fig. \ref{fig_lctype} where
  two periods are shown) of the same models as shown in the Fig.
  \ref{fig_lctype}.
  The phase diagrams were built assuming different values of $\delta_{\phi}$ and
  $\delta_{y}$ that are displayed at the top of each column of panels unless to EA $B*$ and $D*$ panels
  that was assumed a $\delta_{\phi} = 0.025$. Indeed, the eclipse is missed for
  larger $\delta_{\phi}$ values. The regions used to compute the amplitude and
  period uncertainties are indicated by light grey  and dark grey shading in all
  panels  (for more details, see Sects.
  \ref{sec_error} and \ref{sec_testerror}).}
  \label{fig_lcreg}
\end{figure*}

\subsection{Visualizing frequency sampling effects}\label{sec_visualdeltaphi}

Consider a periodic signal of $1d^{-1}$ with measurements covering a variability
cycle from $\mathrm{t = 0}$ to $\mathrm{t = 0 + 1/f}$ and another from 
$\mathrm{t = T_{tot} - 1/f}$ to $\mathrm{t = T_{tot}}$. Five simulated time 
series that mimic pulsating stars ($Y_{(Ceph)}$, $Y_{(RR)}$, $Y_{(RRblz)}$), eclipsing
binary stars ($Y_{(EA)}$), and rotational variables ($Y_{(Rot)}$) were chosen to
illustrate our approach (for more details see Sect.
\ref{sec_testreal}).
Figure \ref{fig_lctype} shows phase diagrams of simulated light curves where 
the first column of panels show the results for $\mathrm{f_{true}}$. The
grey dots indicate the original light curve while the models are indicated by 
purple (Ceph), blue (RR), yellow (RRblz), red (EA), and green (Rot) colours. 
The measurements located at  $\mathrm{t = 0}$ to $\mathrm{t = 0 + 1/f}$ are 
indicated by squares while those at $\mathrm{t = T_{tot} - 1/f}$ to $\mathrm{t
= T_{tot}}$ by crosses. The second, third, and fourth columns show phase
diagrams using $\mathrm{f_{true} + \delta f}$ (see Eq. \ref{eq_deltaf}) for
$\delta_\phi = [0.05,0.1,0.2]$, respectively. The crosses and squares limit the
region where all measurements may be arranged considering that phase values
computed at the beginning and end of the light curve set the largest variation from
the model in the phase diagram as discussed in the Sect.~\ref{sec_freqsamp}. 
As one can see, the largest
distortion of the model is found for binary stars, where the main variation is
concentrated in a small part of the phase diagram. These aspects become
increasingly important in the presence of noise or poorly sampled time series,
when almost all measurements are required to adequately cover all variability 
phases. On the other hand, a low signal-to-noise is found for small frequency 
variations about $\mathrm{f_{true}}$ for those models where the variability is 
observed along the whole phase diagram like Ceph and RR.
Indeed, the phase diagram dispersion is larger for those phenomena whose root
variability causes period and/or amplitude variations like RRLyrae with
the Blazhko effect (RRblz) and rotational variables \citep[e.g.][]{Buchler-2011,
FerreiraLopes-2015cycles}. Indeed, non-radial pulsation, exoplanets, and 
different types of eclipsing and rotational variability enlarge the zoo of phase
diagrams that can be produced by astrophysical phenomena
\citep[e.g.][]{Prsa-2011, DeMedeiros-2013,FerreiraLopes-2015mgiant,
Paz-Chinchon-2015}.
   
To summarize, the phase diagrams of well-defined signals (fixed period and
amplitude) only produce slight variations on the true frequency
and hence these signals are easily identified compared to those ones with
variable period or amplitude where the signal can be completely lost. Of course,
the detection of these stars depends on the susceptibility to each frequency
finding method. These matters will be addressed in a forthcoming paper of this
project. The main conclusion provided by  Eq. \ref{eq_nfmax} is a clear limit to
the variations in which a smooth phase diagram can be found.

\subsubsection{Sorting out $\theta_y$, $\theta_P$, and correction factors}\label{sec_sortreg}

The same models described in the Sect. \ref{sec_visualdeltaphi} were used to
test our assumptions. Figure~\ref{fig_lcreg} shows the phase diagrams of five 
typical light curves where the A panels show the model; the B panels show a
variation in the period with a constant amplitude; constant period with 
amplitude variation (C panels), and both amplitude and period variations (D
panels). These variations were added to the model using a random uniform
distribution, that mimics a non instrumental variation, while an instrumental
variation may appear like a normal distribution. Indeed, the real non
instrumental variation is more complicated and may include variations with
normal, uniform and perhaps more complicated distributions.
For instance, the RR and Rot models at the maximum seem to be composed of 
normal and uniform variations that are not necessarily symmetric about the
model, indicating a more complex variation (see Fig. \ref{fig_lctype} first panels).

Eqs. \ref{eq_neweamp} and \ref{eq_neweperr} can be
considered as a particular case where the noise or variation of amplitude or 
period is provided by a normal distribution since $1.48 \times MAD$ is
approximately the standard deviation value \citep[][]{Hoaglin-1983}. A uniform
distribution has a different spread of values. Therefore, a correction factor
($\gamma$) may be considered in order to take account of the distribution type.
The percentage of values of $68.27\%$, $95.45\%$ and $99.73\%$
that lie within a band around the mean of a normal distribution is
given by $\gamma = 1.48$, $\gamma = 2.96$, and $\gamma = 4.44$, respectively.
However,  $\gamma \simeq 1.37$, $\gamma \simeq 1.92$, and $\gamma \simeq 2.00$ 
contain the same fraction of values if an uniform distribution is considered.
This factor improves our capability to measure an accurate estimation of the
amplitude variation.
For our simulation, this factor is not important since the 
ratio of computed and expected values are analysed (Sect. \ref{sec_testerror}).
On the other hand,  $\gamma = 1.48$ was used to estimate amplitude variation 
on real data (Sects. \ref{sec_testreal} and \ref{sec_catalina}).
The period and amplitude variations computed
are given by the sum of intrinsic and acquired variations. Acquired variations 
are those which come from the environment or instrument while intrinsic
variations come from the source itself. Indeed, low values for the uncertainties
are limited by the instrument properties and for constraints related with 
observability like the sky background, noise from background sources, and
blending. For instance, the period and amplitude variations can reveal 
particularities of phenomena observed if the acquired uncertainties can be 
deducted from a noise model \citep[e.g.][]{Cross-2009,Aigrain-2009,FerreiraLopes-2017papII}. However the
reliability of the period and amplitude variations measured will depend
on the ratio $\delta_{\phi}/\delta_{y}$ as well as the regions used to 
compute them (see Sect. \ref{sec_error} for more detail).

\begin{table}
 \caption[]{Angles ($\theta$) and angular coefficient ($\alpha$) values found 
 for Ceph, RR, RRblz, EA, and Rot models. The angle limits $\theta_y$ and 
 $\theta_P$ with their respective $\alpha_y$ and $\alpha_{\phi}$ values used
 to set the regions to compute the period and amplitude variations as well as
 the maximum angle found in each model are displayed below.} \label{tb_angles}
 \centering    
 \begin{tabular}{l c c c c c}        
 \hline\hline                 
Model & $\mid\theta\mid_{max}$ &  $\theta_y$  & $\mid\alpha_y\mid$ &  $\theta_P$  & $\mid\alpha_{\phi}\mid$ \\    
\hline                         

Ceph & $79.02^o$ & $31.05^o$ & $0.18$ & $76.73^o$ & $4.68$ \\
RR & $85.52^o$ & $28.77^o$ & $0.37$ & $83.67^o$ & $10.99$ \\
RRblz & $81.30^o$ & $18.94^o$ & $0.10$ & $76.25^o$ & $4.74$ \\
EA & $88.39^o$ & $7.87^o$ & $0.04$ & $87.94^o$ & $32.00$ \\
Rot & $74.82^o$ & $34.87^o$ & $0.20$ & $72.90^o$ & $3.48$ \\
\hline                        
 \end{tabular}
\end{table}

\begin{figure}
  \centering
  \includegraphics[width=0.5\textwidth,height=0.8\textwidth]{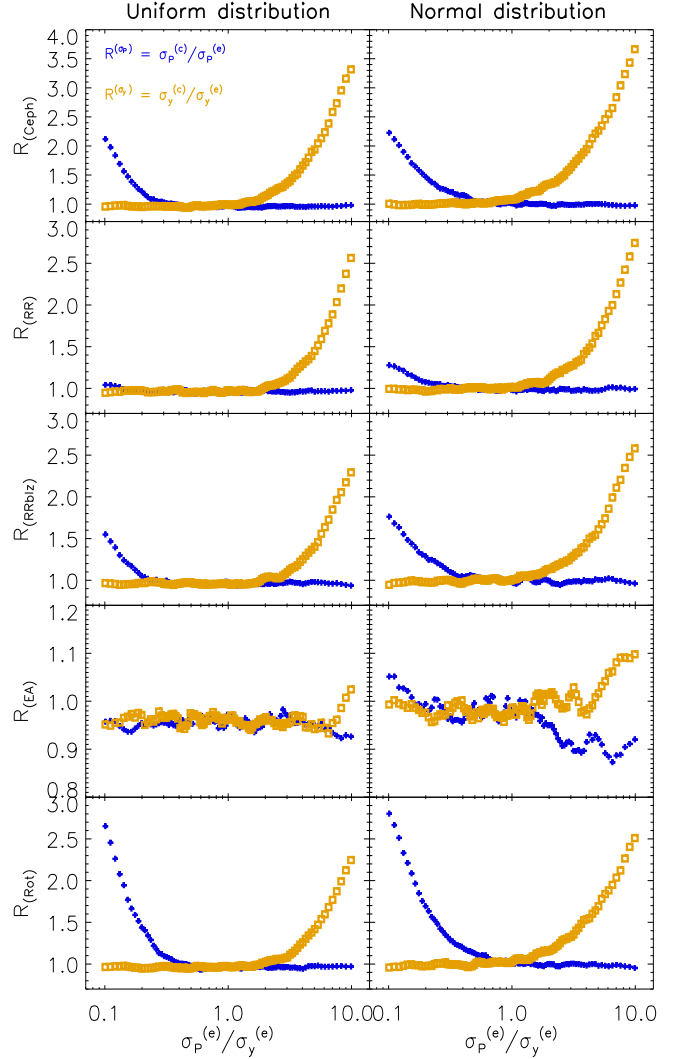} 
  \caption{The ratio of computed and expected uncertainty values for
  period ($R^{(\sigma_{P})}$ - blue crosses) and amplitude ($R^{(\sigma_{y})}$ - yellow
  squares) as function of $\mathrm{\sigma_{P}^{(e)}/\sigma_{y}^{(e)}}$ for  EA,
  RRblz, RR, Ceph, and Rot models. The results where the noise
  was introduced using an uniform and normal distribution are displayed
  in the left and right panels, respectively.}
  \label{fig_simresults}
\end{figure}

\subsection{Testing frequency uncertainities}\label{sec_testerror}

The models described in Sect. \ref{sec_visualdeltaphi} (see A panels of
Fig.~\ref{fig_lctype}) were used to perform the simulations. The regions chosen
to compute the amplitude and period uncertainties are shown in Fig.~\ref{fig_lcreg}. 
The measurements in these regions have angles within the defined angle limits
which were set to best compute the uncertainties (see Sect. \ref{sec_error} for more details). 
Indeed, on average the maximum angle values are reduced and
the minimum angle values are increased when the noise contribution is
increased.
Table~\ref{tb_angles} shows the main parameters values found in each
model. Next, $10^6$ Monte Carlo simulations were performed setting 
$\mathrm{\sigma_{P}^{(e)}/\sigma_{y}^{(e)}}$ in the range from $0.1$ to $10$.
$\mathrm{\sigma_{P}^{(e)}}$ and $\mathrm{\sigma_{y}^{(e)}}$ were introduced 
using a uniform distribution or a normal distribution. Finally,
the amplitude and period uncertainties were computed according to Eqs.
\ref{eq_neweamp} and \ref{eq_neweperr}. The ratio of the computed and expected 
uncertainty values for period ($R^{(\sigma_{P})}$) and amplitude
($R^{(\sigma_{y})}$) were used to estimate the reliability of computed values. 
Figure \ref{fig_simresults} shows the main results obtained in the simulations,
which are summarized below;

\begin{figure}
  \centering
  \includegraphics[width=0.5\textwidth,height=0.6\textwidth]{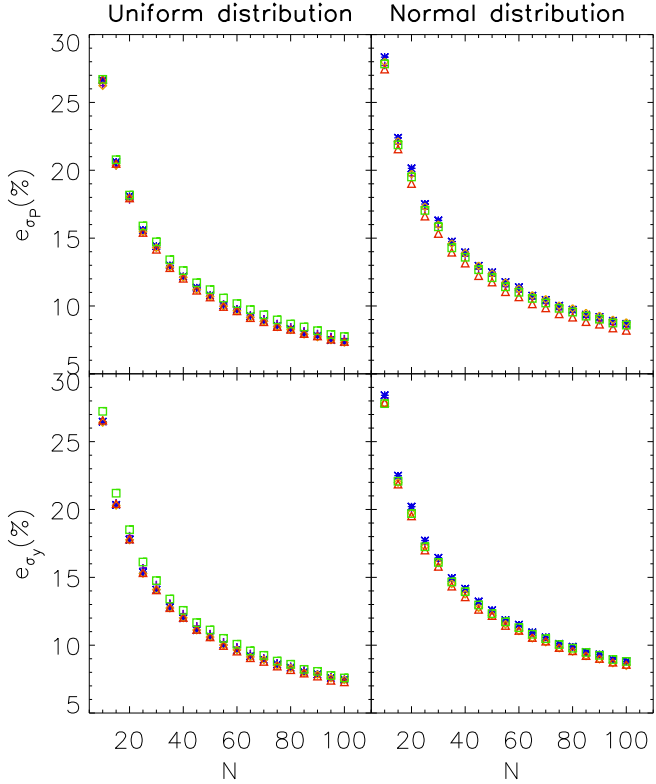} 
  \caption{Percent relative error for  $\sigma_P$ and $\sigma_y$ as a function
  of the number of measurements using an uniform (left panels) and a normal (right panels) distribution.
   The colours indicate the result for different
  models in the same way as Figs. \ref{fig_lctype} and  \ref{fig_lcreg}.}   
  \label{fig_simnumber}
\end{figure}

\begin{itemize}
  \item The results found using uniform and normal
  distributions are quite similar except for EA models. This happens because the
  eclipse is ''missed'' more quickly when the uncertainty is introduced by
  normal distributions than with uniform distributions.
  Considering the same sigma value for both distributions, a normal distribution of errors provides a
  larger dispersion of simulations than an uniform distribution. For instance, 
  $\sim 4.44\times eMAD$ is required to enclose $\sim99.7\%$ of observed
  measurements for a normal distribution while $\sim
  2.00\times eMAD$ is required for a uniform distribution (see Sect. \ref{sec_sortreg}).

  \item $R^{(\sigma_{P})} \simeq  R^{(\sigma_{y})} \simeq 1$ is found for 
  $\mathrm{\sigma_{P}^{(e)}/\sigma_{y}^{(e)}}$ ranging from $\sim0.5$ to $\sim2$ for all models as  well
  as for both uniform and normal distributions. Indeed, the EA
  model has $R^{(\sigma_{P})} \simeq  R^{(\sigma_{y})} \simeq 1$ for almost all
  values of the ratio. $\alpha_{y}$ is smaller than $0.1$ while $\alpha_{\phi}$
  is bigger than $10$ for EA  model and hence the weight of 
  $\mathrm{\sigma_{P}^{(e)}/\sigma_{y}^{(e)}}$ on the computed uncertainties is 
  reduced (see Table \ref{tb_angles}).
   
  \item The greatest difference between computed and expected values (R) are
  found at extreme ratios, i.e those regions where  $\mathrm{\sigma_{P}^{(e)}
  >> \sigma_{y}^{(e)}}$ or $\mathrm{\sigma_{P}^{(e)} << \sigma_{y}^{(e)}}$.
  
\begin{figure*}
  \centering
  \includegraphics[width=0.195\textwidth,height=0.4\textwidth]{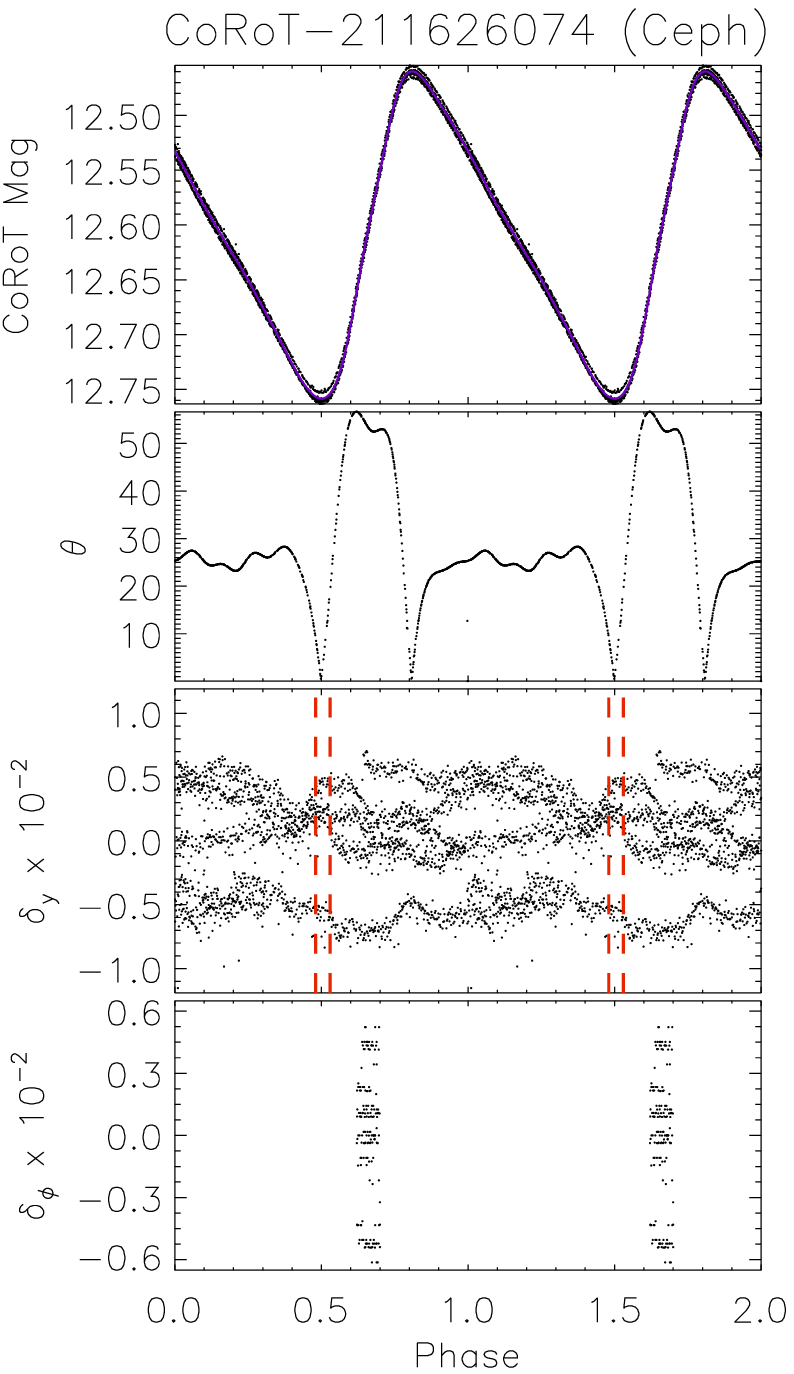} 
  \includegraphics[width=0.195\textwidth,height=0.4\textwidth]{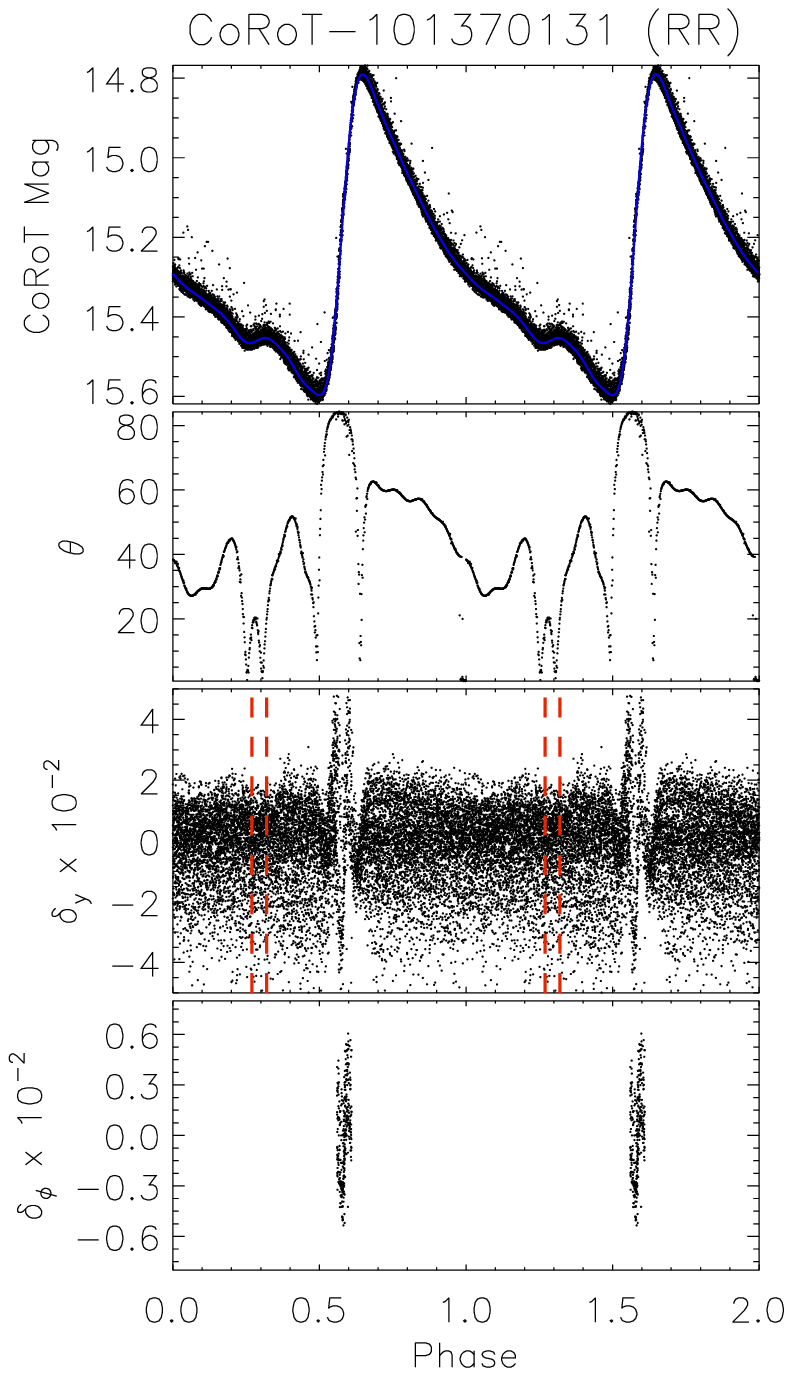} 
  \includegraphics[width=0.195\textwidth,height=0.4\textwidth]{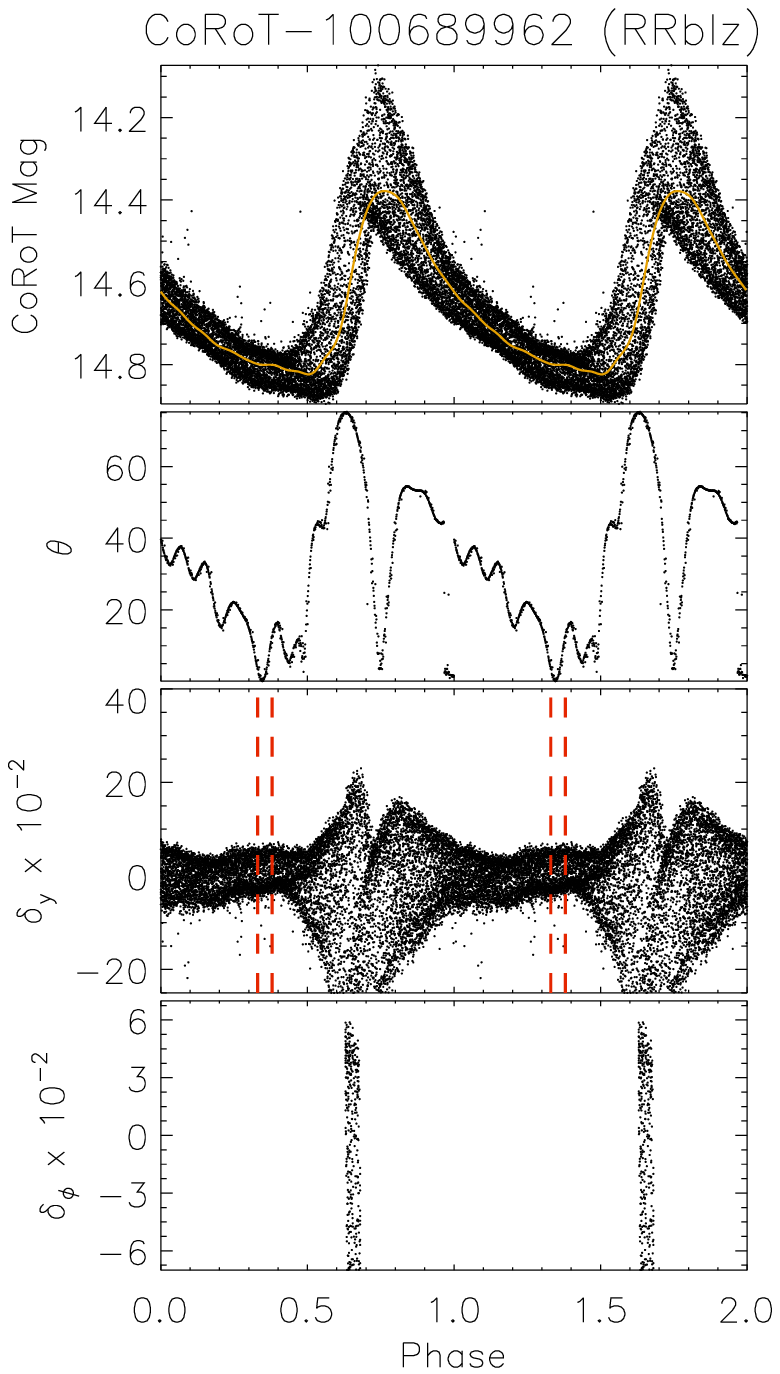} 
  \includegraphics[width=0.195\textwidth,height=0.4\textwidth]{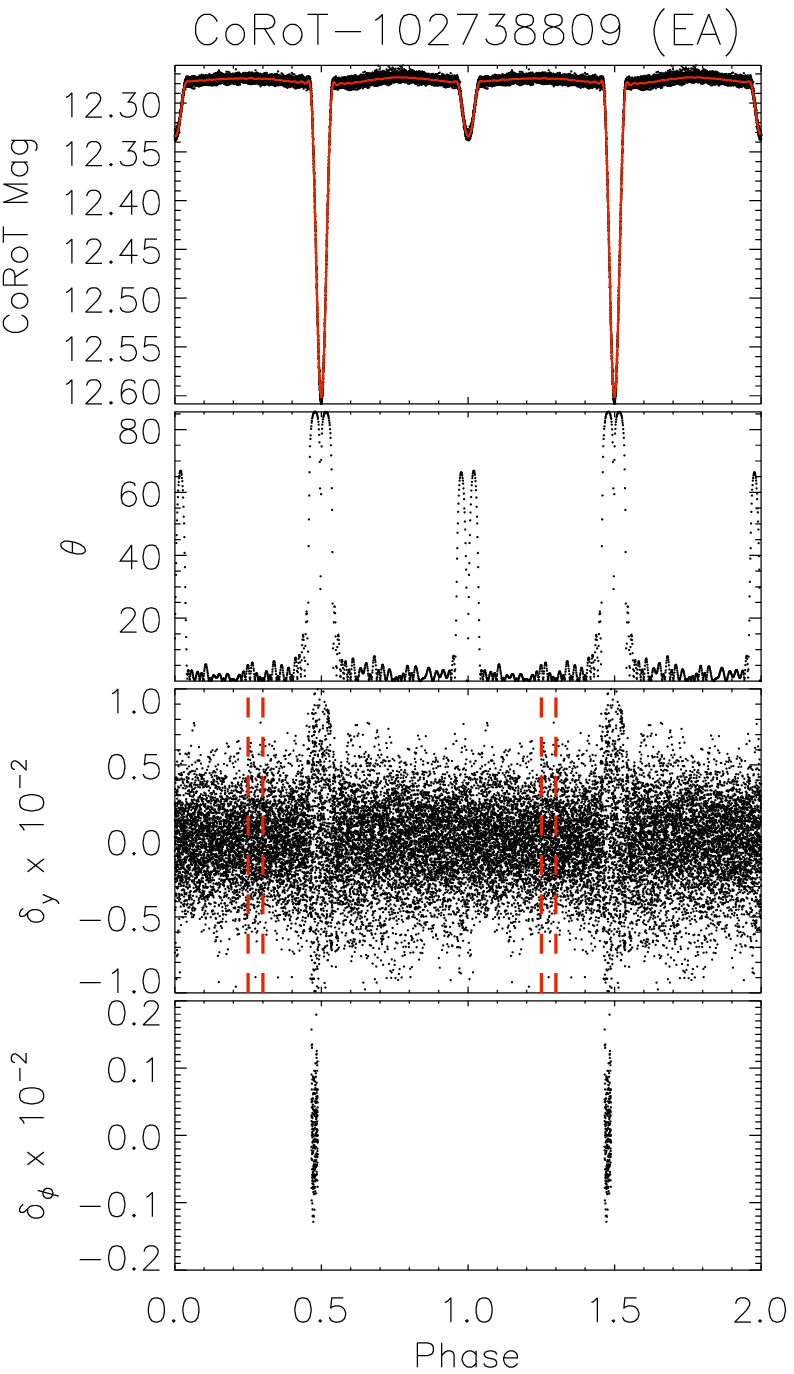} 
  \includegraphics[width=0.195\textwidth,height=0.4\textwidth]{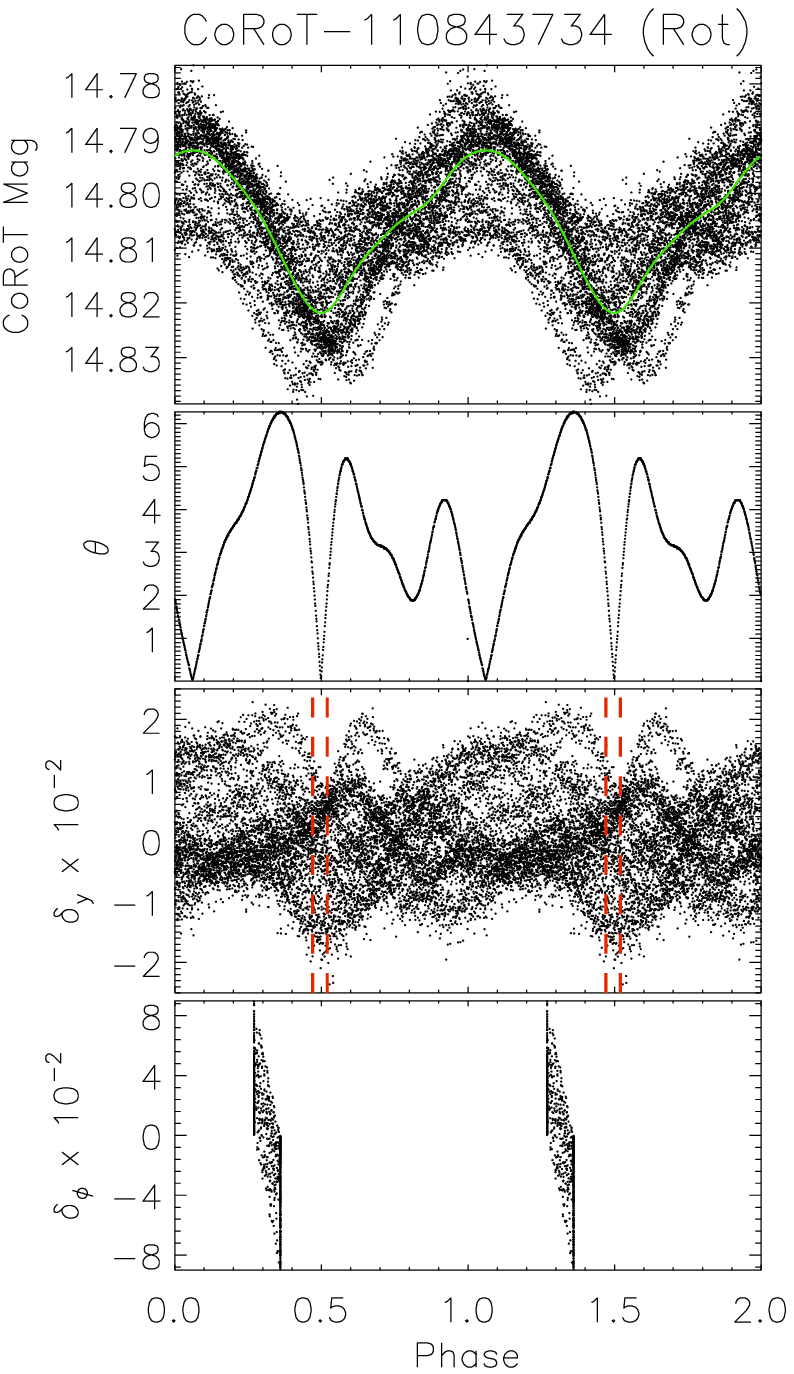} 
  \caption{Ceph (purple - \textit{CoRoT-211626074}), RR (blue - \textit{CoRoT-101370131}),
   RRblz (yellow - \textit{CoRoT-100689962}), EA (red -
   \textit{CoRoT-102738809}), and Rot (green - \textit{CoRoT-110843734}) phase
   diagrams in normalized flux shown in the top row of panels. The angles found
   for each models (second row of panels),  the $\delta_{y}$ (third row of panels),
   and $\delta_{\phi}$ values (bottom row of panels) are also shown.
   Indeed, the last panel only show the results for the region used to compute the period variation. }   
  \label{fig_realdatamodel}
\end{figure*}

\begin{table*}
  \begin{threeparttable}
    
 \caption[]{Parameters for CoRoT stars used to test the approach
 proposed in this work. The L indicate the parameters obtained in the literature
 which the references for are indicated in the last column. Indeed, the
 values of Ra, Dec, R magnitude, and the exposure time ($T_{exp}$) were 
 obtained from the CoRoT database.}\label{tab_modeldata}
 \begin{tabular}{l c c c c c c c c c c c c }        
   \hline\hline                 
  CoRoT-ID & Var. Type &  RA & DEC &  R & $\sigma_{(2h)}$ & $P_{(L)}(d)$ & $\delta P(d)$ & $T_{exp}(d)$ & $A_{(L)}(mag)$  & $eA_{(L)}(mag)$ &  Ref \\   
   \hline
211626074 & Ceph  & 285.469 & 3.277  &  $12.60$ & $1.41\times 10^{-4}$ & $ 5.470600$ & $ \cdot$ & $5.93\times 10^{-3}$ & $ 2.96 \times10^{-1}$ &  $1.44 \times10^{-3}$ &  $ 1$ \\
101370131 & RR    &  292.060 & 0.101 &  $15.28$ & $6.60\times 10^{-4}$ & $6.19332\times10^{-1}$ & $ \cdot$  & $5.93\times 10^{-3}$ & $ \cdot$ &  $ \cdot$  &   $ 2$ \\
100689962 & RRblz &  291.000 & 1.697 &  $14.65$ & $4.60\times 10^{-4}$ & $3.55997\times10^{-1}$ & $ \cdot$  & $5.93\times 10^{-3}$ & $ \cdot$ & $ \cdot$ &   $ 3$ \\
102738809 & EA    &  101.131 & 0.832 &  $12.29$ & $1.18\times 10^{-4}$ & $2.035701$ & $ \cdot$ & $3.70\times 10^{-4}$ & $ \cdot$ & $ \cdot$ &  $ 4$ \\
110843734 & Rot   &  102.918 & -3.748 &  $14.81$ & $5.03\times 10^{-4}$ & $8.186000$ & $ 4.94 \times10^{-2}$ & $5.93\times 10^{-3}$ & $5.66\times10^{-2}$ &  $ 1.42\times10^{-2}$ & $ 5$ \\
 \hline                                   
 \end{tabular}
  \begin{tablenotes}
      \small
    \item The last column is regarding to the references that provided the following parameters above;
    (1) \citet[][]{Poretti-2015}, (2) \citet[][]{Paparo-2009}
    (3) \citet[][]{Chadid-2010}, (4) \citet[][]{Maciel-2011,Carone-2012}
    (5) \citet[][]{DeMedeiros-2013}. Moreover, the noise level ($\sigma_{2h}$) were computed using the Eq. 1 described by \citet[][]{Aigrain-2009} where z was computes as the mean value of CoRoT Run analysed by the authors.      
    \end{tablenotes}
  \end{threeparttable}
\end{table*}

  \item The discussion above can be summarized if the Eqs. \ref{eq_uncamprel}
  and \ref{eq_unperrel} are extrapolated thus:
  
   \begin{equation}
    R^{(\sigma_{y})} = \frac{\sigma_{y}^{(c)}}{\sigma_{y}^{(e)}} \simeq \left|\alpha_{y}\right| \times \frac{\sigma_{P}^{(e)}}{\sigma_{y}^{(e)}} + 1
   \label{eq_uncamprel2}    
  \end{equation} 
 
  and
 
   \begin{equation}
       R^{(\sigma_{P})} =\frac{\sigma_{P}^{(c)}}{\sigma_{P}^{(e)}} \simeq \left|\alpha_{\phi}^{-1}\right| \times \left(\frac{\sigma_{P}^{(e)}}{\sigma_{y}^{(e)}} \right)^{-1} + 1.
   \label{eq_unperrel2}    
  \end{equation}

$R^{(\sigma_{y})}$ and $R^{(\sigma_{P})}$have opposite behaviour since
they vary with $\mathrm{(\sigma_{P}^{(e)}/\sigma_{y}^{(e)}})^{\pm1}$,
respectively. $R^{(\sigma_{y})}$ implies a rational
function if $\mathrm{\sigma_{P}^{(e)}/\sigma_{y}^{(e)}}$ has values
smaller than $1$ while the opposite is found for $R^{(\sigma_{P})}$. 
However, both functions depend on an angular coefficient
($\left|\alpha_{y}\right|$ or $\left|\alpha_{P}\right|$) that will 
determine the trend variation.
\end{itemize}

The simulations are in agreement with the analysis in Sect. \ref{sec_error}.
The amplitude and period variations can bias the uncertainty estimations of one
another, mainly when $\mathrm{\sigma_{P}^{(e)}/\sigma_{y}^{(e)}} << 1$ or 
$\mathrm{\sigma_{P}^{(e)}/\sigma_{y}^{(e)}} >> 1$. Indeed, Eqs.
\ref{eq_uncamprel2} and \ref{eq_unperrel2} can be used to estimate the 
reliability of uncertainties if  $\mathrm{\sigma_{P}^{(e)}/\sigma_{y}^{(e)}}$
can be estimated somehow.

The relative errors of the uncertainties were also analysed as function
of the number of measurements (see Fig. \ref{fig_simnumber}). As result, a decrease in
the error with the number of measurements is found, as expected. This means that
the number of measurements is an implicit parameter in Eqs. \ref{eq_neweamp}
and \ref{eq_neweperr} that improve the statistic significance of uncertainties.

\subsection{Describing Models and Testing the Approach on Observed
Data}\label{sec_testreal}

Ceph, RR, RRblz, EA, and Rot models were based on the 
CoRoT light curves \textit{CoRoT-211626074}, \textit{CoRoT-101370131}, 
\textit{CoRoT-100689962}, \textit{CoRoT-102738809}, and 
\textit{CoRoT-110843734}, respectively. The variability types  were previously identified
by
\citet[][]{Debosscher-2007,Poretti-2015,Paparo-2009,Chadid-2010,Maciel-2011,Carone-2012}
and \cite[][]{DeMedeiros-2013}. Table \ref{tab_modeldata} shows the main
parameters of these sources that were obtained in the literature (L). These
light curves were  modeled using a harmonic fit  with  $12$, $12$, $12$, $24$, and $4$
harmonics for $Ceph$, $RR$, $RRblz$, $EA$, and $Rot$ variable stars, respectively.
Higher number of harmonics can be used, however this also increases
the processing time necessary to model and to perform simulations. The
$Y_{(RRblz)}$ and $Y_{(Rot)}$ variable stars present variations in the
amplitude and a period-amplitude variation. The $Y_{(RRblz)}$ has a
Blazhko effect that is a long-period modulation or a variation in period and
amplitude of RR Lyrae stars \citep[e.g.][]{Szabo-2014}. On the other hand, the 
$Y_{(Rot)}$ displays amplitude variation due the to the magnetic activity cycles
and period variation due to differential rotation
\citep[e.g.][]{FerreiraLopes-2015cycles,DasChagas-2016}.
The exposure time ($T_{exp}$) provided by CoRoT mission and  the
empirical noise relation ($\sigma_{2h}$) described by \citet[][]{Aigrain-2009}
were  used to analysis the period and amplitude variation.

The tests performed in the sections \ref{sec_error}, 
\ref{sec_visualdeltaphi}, and \ref{sec_sortreg} used models scaled to unit amplitude.
It is useful to test our approaches for different signal types.
For instance, the Ceph, RR, RRblz, EA, EB and Rot models have similar
angles (see Table \ref{tb_angles}) but a wide difference among them is found
when the real data is considered (see Fig. \ref{fig_realdatamodel})
since they have different typical amplitudes. Therefore, the angles found in the
real data are not the same as those found for the models
tested in the previous sections, as expected.
These variations occur because $\tan(\theta) = \delta y/ \delta \phi $, i.e.
a bigger $\delta y$ for the same $\delta \phi$ implies a larger angle.
Figure \ref{fig_realdatamodel} shows the CoRoT light curves (first row of panels),
the angles as a function of phase along the light curve (second row of
panels), the observed minus modeled values (third row of panels),
and finally the $\delta \phi$ values for the region used do compute the period variation.
For example, the $\theta_{max}$ for $Rot$ models is about twelve times bigger 
than those found when amplitude is scaled to unit amplitude.
On the other hand, the $\theta_{max}$ of the Ceph, RR, RRblz, EA, and EB 
decrease by factors smaller than $0.5$.

\begin{figure*}
  \centering
   \includegraphics[width=0.95\textwidth,height=0.3\textwidth]{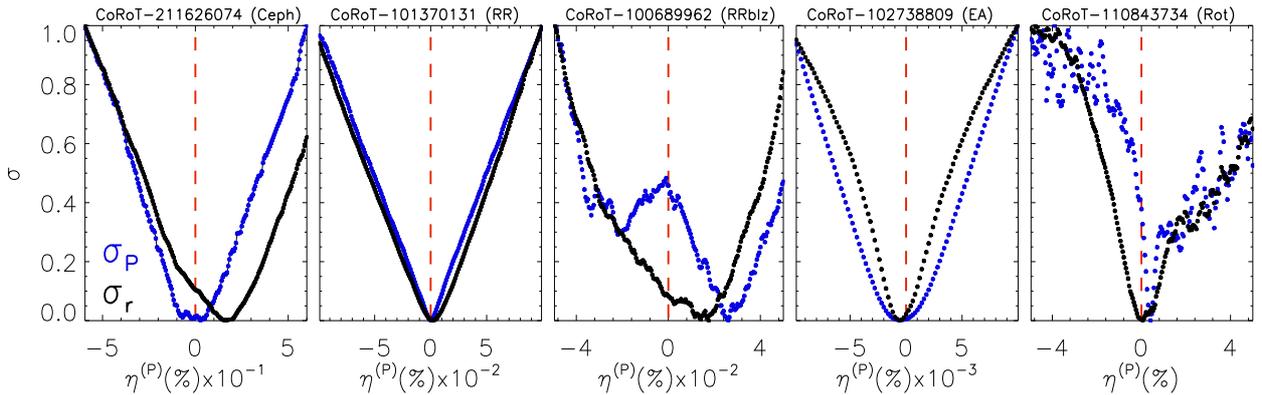} 
  \caption{The normalized standard deviation scaled between 0 and 1 as 
  function of the percent relative error of $P_L$.
  The results for the residuals and period are shown as black and blue
  dots, respectively. The red line sets the location of the variability
  period determined in literature $P_{(L)}$.  }
  \label{fig_normalizedsigma}
\end{figure*}

\begin{table*}
 \caption[]{Parameters for CoRoT stars computed from the aproaches proposed in this work.}\label{tab_modeldatanew} 
 \begin{tabular}{l c c c c c c c c c}        
   \hline\hline                 
  CoRoT-ID &  $P(d)$  &  $\delta  P_{(FWHM)}^{(LSG)}(d)$ &  $\delta P_{(FWHM)}^{(STR)}(d)$ & $\sigma_P(d)$ & $A(mag)$ & $\sigma_A(mag)$ & $\sigma_r(mag)$  & $\sigma_P/\sigma_A$ & $Log(T_{tot}/P)$  \\   
   \hline
 211626074  & $5.47073174$  & $4.55\times 10^{-1}$  & $3.33\times 10^{-1}$  & $5.27\times 10^{-3}$  & $2.99\times 10^{-1}$  & $1.58\times 10^{-3}$  & $1.87\times 10^{-3}$ & $2.86$ & $6.69\times 10^{-1}$  \\
101370131  & $6.19331408\times 10^{-1}$  & $9.15\times 10^{-4}$  & $7.46\times 10^{-4}$  & $5.80\times 10^{-4}$  & $7.98\times 10^{-1}$  & $7.53\times 10^{-3}$  & $8.14\times 10^{-3}$  & $0.06$  & $2.39$  \\
100689962  & $3.56090879\times 10^{-1}$  & $3.27\times 10^{-4}$  & $2.93\times 10^{-4}$  & $2.23\times 10^{-2}$  & $4.50\times 10^{-1}$  & $3.82\times 10^{-2}$  & $6.19\times 10^{-2}$   & $0.52$ & $2.60$  \\
102738809  & $2.03569293$  & $1.17\times 10^{-2}$  & $3.02\times 10^{-3}$  & $7.02\times 10^{-4}$  & $3.30\times 10^{-1}$  & $3.47\times 10^{-3}$  & $3.52\times 10^{-3}$ & $0.25$  & $1.81$  \\
110843734  & $8.21895695$  & $1.74\times 10^{-1}$  & $1.70\times 10^{-1}$  & $1.71\times 10^{-1}$  & $2.68\times 10^{-2}$  & $6.06\times 10^{-3}$  & $7.76\times 10^{-3}$  & $30.01$  & $1.13$\\
   \hline
 \end{tabular}
\end{table*}

Figure \ref{fig_realdatamodel} displays, step by step, the procedure
that must be used to compute period and amplitude variations: the variability period
is computed and the light curve is folded; next, a model is obtained using 
harmonic fits (see solid lines in the upper panels); from the models the angles 
are determined (see second row of panels) from where the regions used to compute
period and amplitude variations are established; the amplitude variation is
given by the standard deviation of the residuals in the region of phase
diagram where $\left| \theta \right| < \theta_y$;  and the period variation 
is found by multiplying the variability period by $eMAD$ of $\delta_{\phi}^{(c)}$ 
(given by Eq. \ref{eq_neweperr}).
The periods and amplitudes as well as their uncertainties and variations were
computed as described in Sect. \ref{sec_computeerror}  (see Table \ref{tab_modeldatanew}).
The results were compared with previous ones 
(see Table \ref{tab_modeldata}) where the main remarks are summarized below;

\begin{itemize}
   \item The period that leads  to the smallest $\sigma_{P}$ is not always 
   related with the smallest $\sigma_{r}$ (see Fig. \ref{fig_normalizedsigma}).
   
   \item All $\sigma_{A}$ values are bigger than those
   given by $\sigma_{2h}$. This indicates an underestimation of
   $\sigma_{2h}$ or that all sources have an intrinsic amplitude variation.
   A note of caution, the noise values decrease with $Log(T_{tot}/P)$ and hence
   such a comparison cannot be performed straightforwardly.
       
   \item \textit{CoRoT-100689962},
   \textit{CoRoT-110843734}, and \textit{CoRoT-102738809} have $\sigma_{P}$ 
   values larger than the exposure time ($T_{exp}$). However,
   \textit{CoRoT-101370131} has  $\sigma_{P}$ ten times smaller than $T_{exp}$.
    
   \item The $\sigma_{P}/\sigma_{A}$ values for all 
   sources are smaller than $\sim0.5$ or bigger than $2$ (see Sect.
   \ref{sec_testerror}).
   It indicates that all  $\sigma_{P}$ and $\sigma_{A}$ values 
   are biased by amplitude or period variation, respectively.
   Indeed, the intrinsic variation is not known a priori and hence the
   information provided by the ratio $\sigma_{P}/\sigma_{A}$ will only be accurate if
   $\mid\alpha_y\mid << 0$ and $\mid\alpha_{\phi}\mid >> 1$ (see Sect. \ref{sec_error}).
        
   \item The variability periods determined by us are in agreement with
   those found in the literature. Indeed, the literature periods are determined
   as the highest power spectrum peak  while those found by us are
   calculated by minimising $\sigma_{P}$.   
   
   \item The period uncertainty $\delta P_{(FWHM)}^{(STR)}$ method is
   always smaller than $\delta P_{(FWHM)}^{(LSG)}$ that indicates that STR
   is more sensitive to variation in the phase diagram than the LSG method.
   
   \item \textit{CoRoT-211626074} -  The amplitude 
   ($A_{(L)}$) found in the literature is about $1\%$ smaller than that found 
   by us. However, the authors used the DR2 release while our data come from
   the DR4 release.
   Indeed, the amplitudes are in agreement within the error bars.
   The $\sigma_{A}$ is at least nine times bigger than  $\sigma_{2h}$. 
   Moreover, $\sigma_{P}/\sigma_{A} = 2.86$ indicates
   that the weight of $\sigma_{P}$ in $\sigma_{A}$ is not strong, and vice-versa.
   It indicates that some of the  amplitude variation comes from the sources.
   This result is supported by the detection of overtone pulsation reported by 
   \citet[][]{Poretti-2015}.
   Indeed, the determination of amplitude variation reported by us was only
   settled by determination of $\sigma_{A}$ while the authors use complex
   analysis.
   
   \item \textit{CoRoT-101370131} -  The $\sigma_{P}$ 
   is smaller than $T_{exp}$ indicating a non-intrinsic variation related with 
   the period. On the other hand, the amplitude variation $\sigma_{A}$ is about
   nine times bigger than $\sigma_{2h}$. Moreover,  $\sigma_{P}/\sigma_{A} =
   0.06$ also indicates that $\sigma_{A}$ is not biased by $\sigma_{P}$.
   Therefore, an intrinsic variation of the \textit{CoRoT-101370131} can be real
   if the noise level estimation is reliable.
      
   \item \textit{CoRoT-100689962} - The period and amplitude variation
   is clearly observed in the phase diagram.
   Moreover, it has the largest $Log(T_{tot}/P)$ and hence the 
   smallest $\delta P_{(FWHM)}$ in agreement with the discussion performed in
   Sect. \ref{sec_freqerror}. Moreover, $\sigma_{P}/\sigma_{A} = 0.52$ indicates
   that the period variation is not strongly biased by amplitude variation and vice-versa.    
   Therefore, the $\sigma_{P}$ and $\sigma_{A}$ mean that
   intrinsic variations come from the source since these variations are
   $\sim3.8$ times bigger than $T_{exp}$ and $\sim12$ times bigger than 
   $\sigma_{2h}$, respectively.
   
   \item \textit{CoRoT-102738809} - The $\sigma_{P}$ is 
   the smallest value among the sources analysed. Indeed, this aspect is
   caused by the large angles and the shape of the light curve. Moreover, this
   source has the shortest exposure time (see Table 
   \ref{tab_modeldata}).
   The $\sigma_{P}$ does not show strong evidence of a period variation since it
   is smaller than twice $T_{exp}$. On the other hand, $\sigma_{A}$ is three times
   larger than $\sigma_{2h}$ that indicates a small intrinsic variation related
   with the amplitude. Indeed, the region used to compute the amplitude variation is 
   related with the eclipse phase where both stars are side by side. Therefore, 
   $\sigma_{A}$ can be related to one or both stars.

   \item \textit{CoRoT-110843734} - The $\delta
   P_{(FWHM)}$ is bigger than $\delta P$ and hence  the empirical relation given
   by \citet[][]{Lamm-2004} can provide values smaller than those found for the 
   $\delta P_{(FWHM)}$ estimations. $A_{(L)}$ is about twice that estimated by
   us. Such a difference can only be achieved by a typing error.
   On the other hand, the period computed by the authors is in agreement with 
   that found for us. $\theta_{max} \sim 6^{o}$ and hence the period variation
   is biased by amplitude variation. Indeed, $\sigma_{P}/\sigma_{A} = 30.0$ indicates
   an unreliable estimation of $\sigma_{P}$ using the phase diagram.
   Therefore, $\sigma_{P}$ or $\sigma_{A}$ are not useful as indicators of
   intrinsic variation for rotational variables having small amplitudes.
   However, the estimation of period and amplitude variation with time instead
   of phase can reveal important clues about stellar activity cycles
    \citep[e.g.][]{FerreiraLopes-2015cycles}.  
\end{itemize}

In summary, the period and amplitude variation can provide important
information about the intrinsic variation of the source. However, it is
trustworthy only if $\theta_{max} >> \theta_{min}$ since the capability to
discriminate period and amplitude variation decreases. For a complete
characterization of a light curve the period uncertainty as well as period 
variation must be determined.

\begin{figure*}
  \centering
   \includegraphics[width=0.48\textwidth,height=0.3\textwidth]{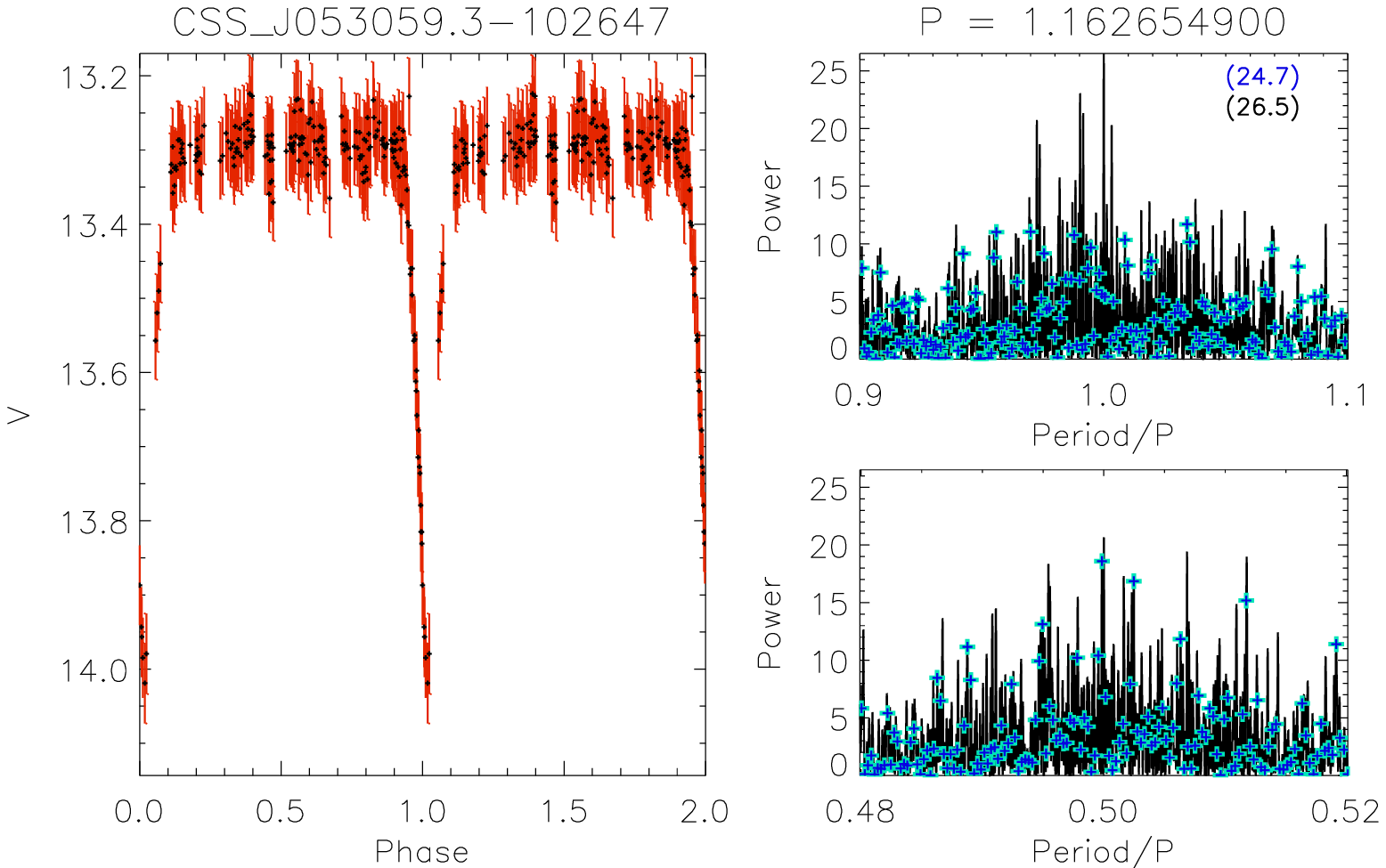} 
   \includegraphics[width=0.48\textwidth,height=0.3\textwidth]{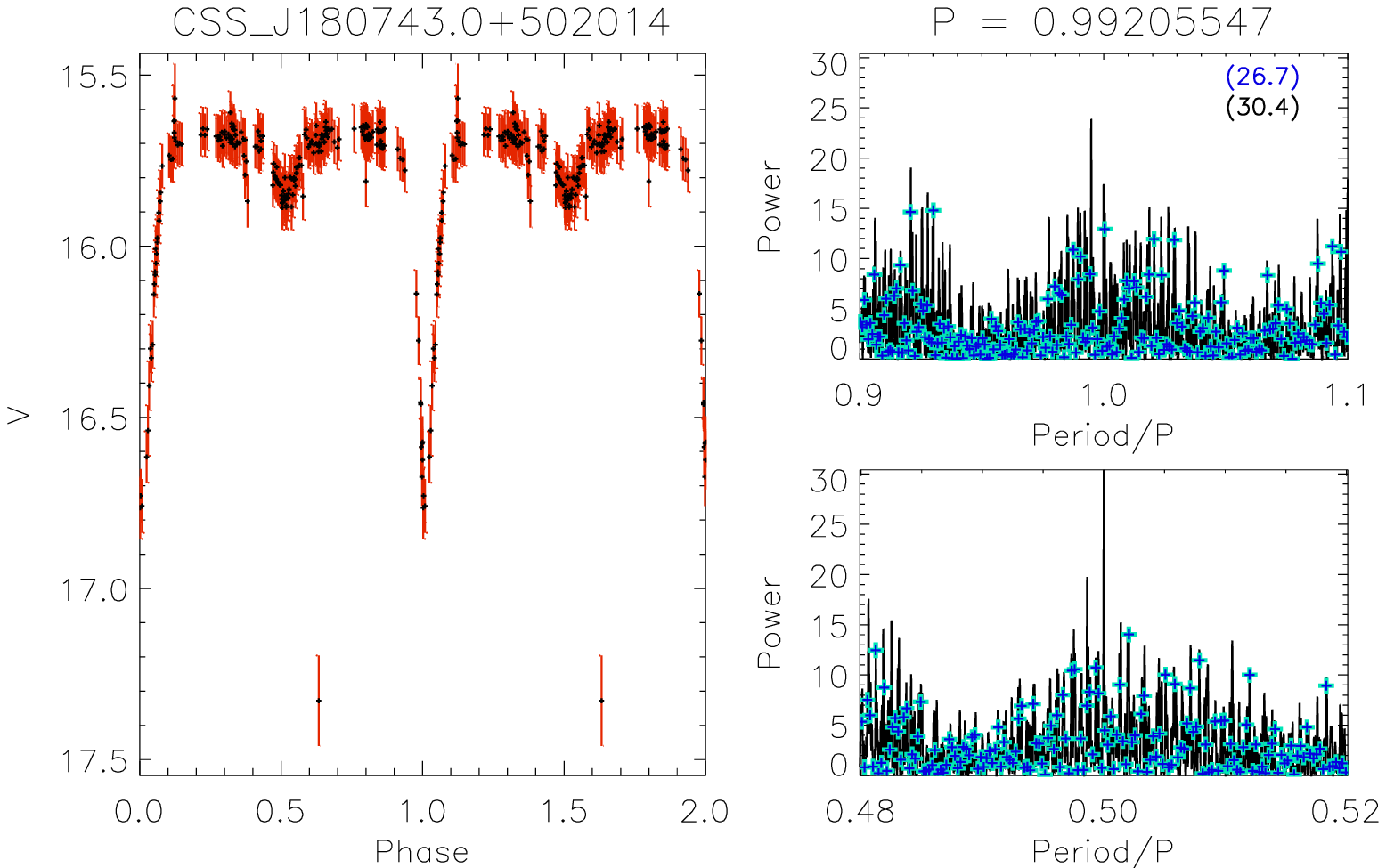} 
   \includegraphics[width=0.48\textwidth,height=0.3\textwidth]{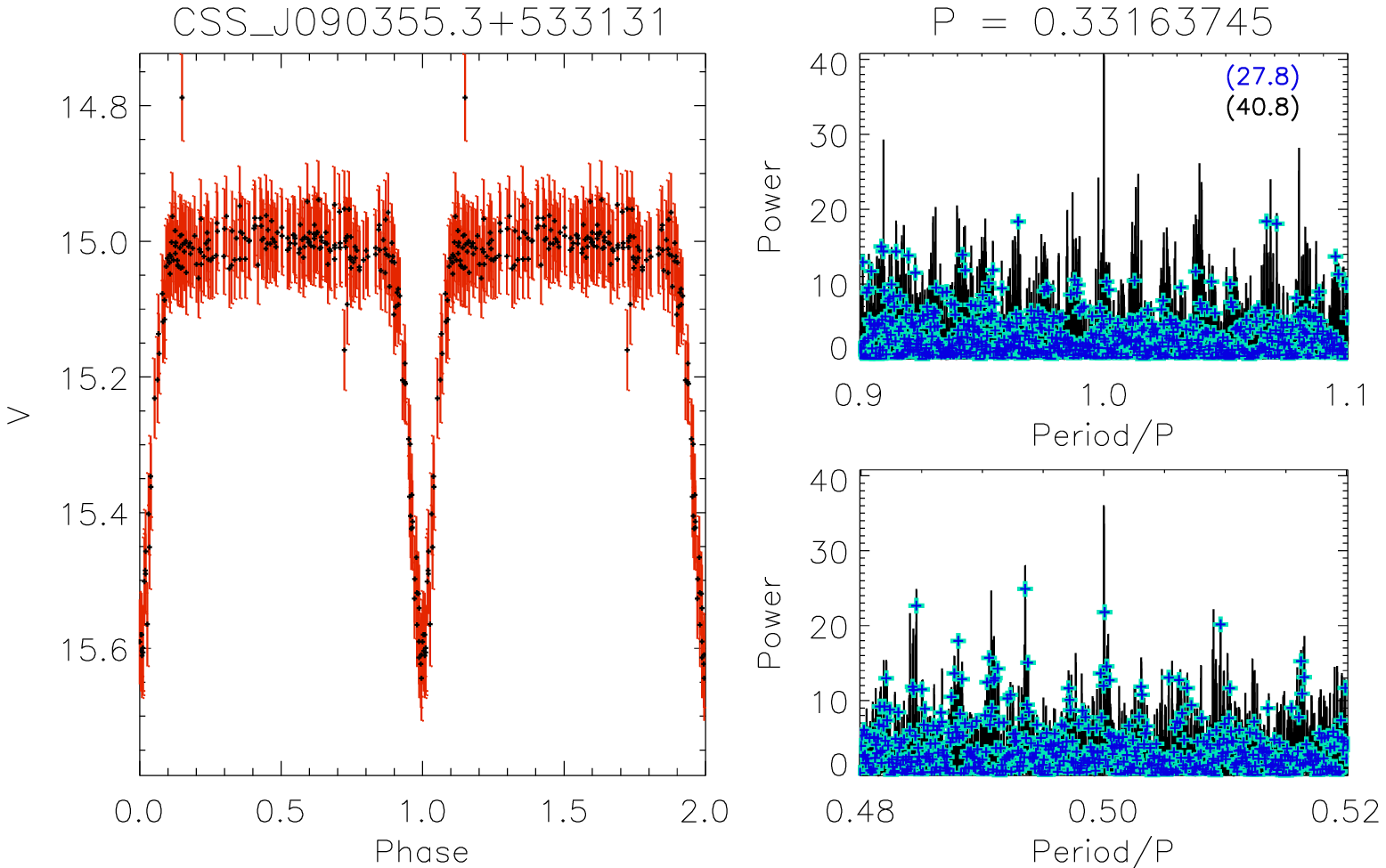} 
   \includegraphics[width=0.48\textwidth,height=0.3\textwidth]{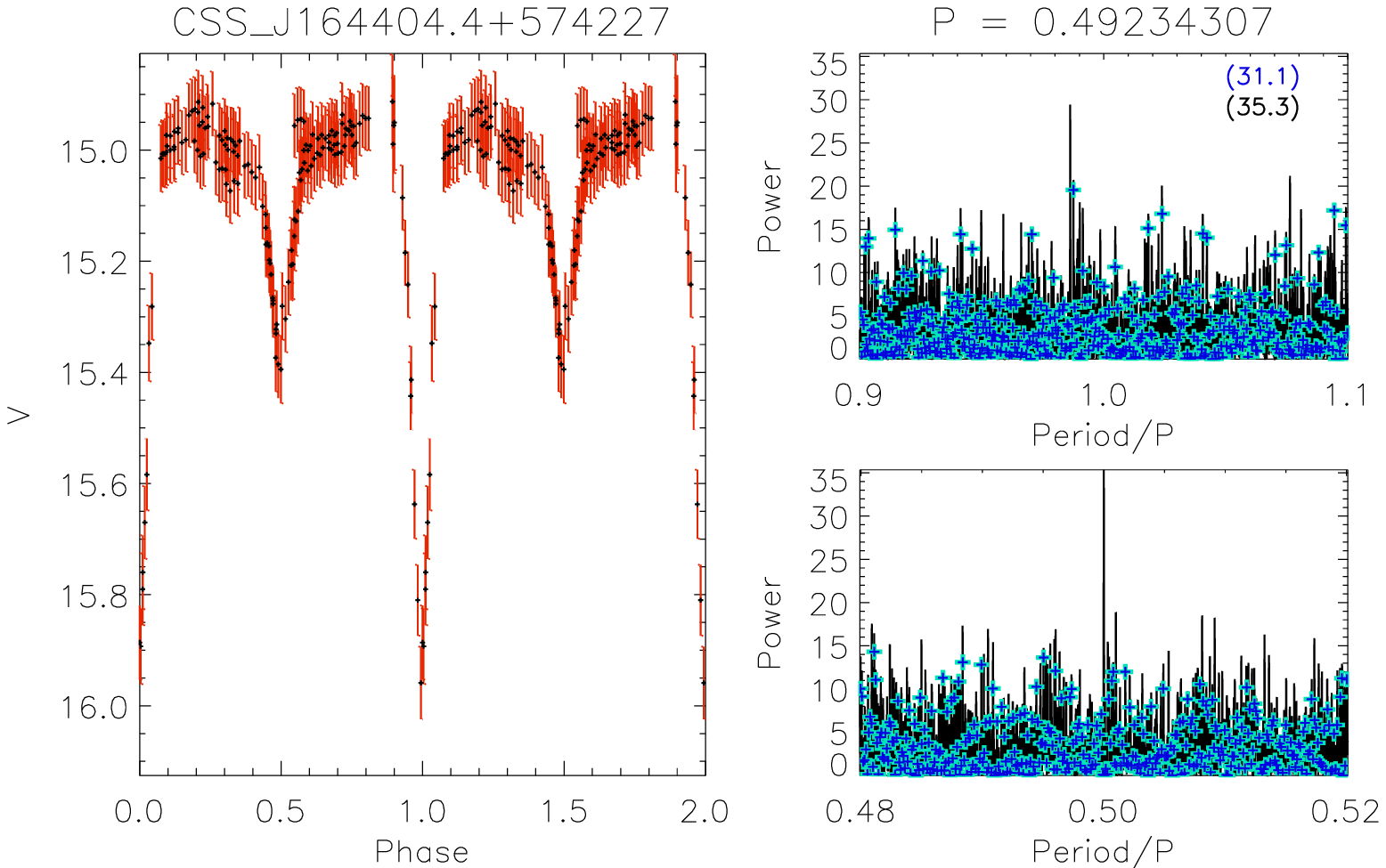} 
   
  \caption{ Phase diagram of 4 $EA_{up}$ Catalina stars
  where we also display the Lomb-Scargle periodogram about the 
  variability period (top right panel for each star) and about half the
  variability period (bottom right panel for each star).
  The name and the variability period is shown in each diagram. The solid black
  line shows the power spectrum considering a number of frequencies obtained
  from  Eq.\ref{eq_nfmax} assuming $\delta_{\phi} = 0.01$ while those values 
  found using $N_{freq} = 10^4$ are marked by blue crosses.
  Moreover, the maximum power found for both methods of frequency sampling is
  displayed in the upper right corner. }
  \label{fig_eacatalina}
\end{figure*}

\subsection{Testing Frequency Sampling on Observed Data}\label{sec_catalina}

The Catalina Real Time Survey found about $\sim47000$
periodic variables stars in Data Release-1 \citep[][]{Drake-2014}. The authors
reported a sample of EA variables stars where the period determination was not
possible due an insufficient number of observations at the eclipses. These stars
were reported as EA variables having unknown-period ($EA_{up}$). 
 The Lomb-Scargle method \citep[][]{Lomb-1976,Scargle-1982} was
used to perform a period search but the frequency range and frequency sampling
are not given by the authors.
Therefore, a mean value of those shown in the Table  \ref{tabconstraint} were assumed,
i.e. $F_{min} = 2/T_{tot}$,  $F_{max} = 10$, and $N_{freq} = 10^4$.
These constraints were assumed as those used by the authors to review a small
sample of $EA_{up}$ stars.

Indeed, EA stars require a high frequency sampling to
allow  us to determine the variability period otherwise the eclipse region will 
not be smoothly folded (see $Y_{(EA)}$ panel Fig. \ref{fig_lctype}).
Section \ref{sec_freqsamp} discussed the frequency sampling in detail where 
the $\delta_{\phi}$ required to find the variability periods for EA stars is
smaller than $0.05$ in order to be able to fold the eclipse properly.
Therefore, four $EA_{up}$ Catalina stars (see Table \ref{table_eacatalina}) were
reviewed using the frequency sampling given by $\delta_{\phi} = 0.01$. Indeed,
the sample analysed has $T_{tot} \simeq 3000$ days that
implies a number of frequencies $\sim 3\times10^6$ (see Eq.\ref{eq_nfmax}).

Figure \ref{fig_eacatalina} shows four $EA_{up}$ stars
where the variability period was determined. In the right panel of each phase
diagram is shown the Lomb-Scargle power spectrum about the variability period 
using $N_{freq} = 10^4$ (blue crosses) and a number of frequencies obtained from
Eq.\ref{eq_nfmax} assuming $\delta_{\phi} = 0.01$ and $F_{max} = 10$. As one can
verify the highest peak of the black lines is related to the maximum power of the
periodogram.
On the other hand, these peaks are not found when the sampling frequency is
reduced (blue crosses). Therefore, the  variability periods of $EA_{up}$ stars
were not identified due to low frequency sampling. The main parameters of the
four $EA_{up}$ stars analysed in the present work are presented in the Table
\ref{table_eacatalina}.

\begin{table*}
 \caption[]{ Parameters for $EA_{up}$ Catalina stars computed from the
 approaches proposed in this work.}\label{table_eacatalina}
 \scriptsize
 \begin{tabular}{l c c c c c c c c c}        
   \hline\hline                 
Catalina-ID &  $P(d)$  &  $\delta  P_{(FWHM)}^{(LSG)}(d)$ &  $\delta  P_{(FWHM)}^{(STR)}(d)$ & $\sigma_P$ & V & $A$ & $\sigma_A$ & $\sigma_R$  & $Log(T_{tot}/P)$ \\   
  \hline       
CSS\_J053059.3-102647 & $1.16265491$ & $1.32\times 10^{-4}$ & $7.69\times 10^{-5}$ & $2.08\times 10^{-3}$ & $13.350$ & $7.50\times 10^{-1}$ & $2.83\times 10^{-2}$ & $2.17\times 10^{-2}$ & $3.40$ \\
CSS\_J180743.0+502014 & $9.92055466\times 10^{-1}$ & $8.20\times 10^{-5}$ & $6.99\times 10^{-5}$ & $1.22\times 10^{-3}$ & $15.810$ & $1.14$ & $2.71\times 10^{-2}$ & $1.09\times 10^{-1}$ & $3.49$ \\
CSS\_J090355.3+533131 & $3.31637448\times 10^{-1}$ & $1.33\times 10^{-5}$ & $2.49\times 10^{-5}$ & $2.27\times 10^{-3}$ & $15.095$ & $6.32\times 10^{-1}$ & $2.06\times 10^{-2}$ & $3.01\times 10^{-2}$ & $3.94$ \\
CSS\_J164404.4+574227 & $4.92343069\times 10^{-1}$ & $1.14\times 10^{-5}$ & $4.56\times 10^{-5}$ & $1.18\times 10^{-3}$ & $15.088$ & $9.92\times 10^{-1}$ & $1.64\times 10^{-2}$ & $2.91\times 10^{-2}$ & $3.74$ \\
  \hline  
 \end{tabular}
\end{table*}

Indeed, from a methodology viewpoint, the identification of variability periods
of $EA_{up}$ stars requires a suitable period finding method and high frequency 
sampling (see Sect. \ref{sec_freqsamp}) to detect the signal.
Moreover, the susceptibility of the period finding methods varies for different
signal shapes (see Fig. \ref{fig_lctype}).
Therefore, a deeper analysis of all $EA_{up}$ stars will be performed in
a forthcoming paper where other methods besides Lomb-Scargle  will also be used.
As a result, we will define limits on what constitutes an insufficient number
of observations.

\section{Conclusions}\label{sec_conclusion}

Frequency analysis constraints as well as the period and amplitude variations
were analysed in the present work. A new approach to compute frequency sampling
was introduced. This analysis is fundamental to providing a precise number of
frequencies required to perform period finding searches. It also enables us
to identify optimal values for searching for particular variable types
as well as how much resolution is required to increase the accuracy of the periods found.
We consider that this approach is fundamental to efficiently face the challenges
of big data science since analytical equations are imposed. 

The period and amplitude variation of light curves were also reviewed. 
We show that a complete characterization of a light curve requires separating
period uncertainty and period variation, from which important information about
the variability nature can be estimated. On the other hand, the noise and
amplitude variation also provide new clues about intrinsic variations that come from
the source. The analyses performed in this project are very
useful since all aspects of the analyses of large photometric surveys are being
studied in order to maximize the probability of finding variable stars, reduce 
the running time, and reduce the number of misclassifications. The current paper
is the second step towards unbiased samples, i.e. samples that only enclose 
reliable variations since this is unfeasible using correlated or non-correlated 
indices alone. Moreover, in this project we try to standardize the analysis 
criteria for variable stars in photometric surveys. In spite of this, the 
dependence of variability indices on the instrumental properties has been
reduced and now, we also propose an estimation of frequency sampling that 
reduces the dependence on the total time span or time sampling of the data. 
Moreover, an approach to study the amplitude and period variation is 
presented. We consider that these estimations provide better information  about
the phenomena observed than previous ones since these estimations are limited by 
instrument properties or signal features. These must be taken into account for a
realistic estimation.

This paper concludes our studies about the constraints used to perform
frequency searches. A new frequency finding method and new insights to detect
aperiodic variations and to determine the false alarm probability will be
addressed in a forthcoming paper of this series.

\section*{Acknowledgements}
C. E. F. L. acknowledges a post-doctoral fellowship from the CNPq.
N. J. G. C. acknowledges support from the UK Science and Technology Facilities Council.
The authors thank to MCTIC/FINEP (CT-INFRA grant 0112052700) and the Embrace Space Weather Program for the computing facilities at INPE.

\bibliographystyle{mnras} 
\bibliography{mylib_1611.bib}

\bsp	
\label{lastpage}
\end{document}